\documentclass[12pt]{article}
\usepackage[]{graphics}
\begin{document}
\begin{flushleft}
{\it Yukawa Institute for Theoretical Physics}
\begin{flushright}
YITP-95-19\\
December 20, 1995\\
hep-th/9512158
\end{flushright}
\end{flushleft}

\begin{center}
\vspace*{1.0cm}

{\LARGE{\bf  Non-canonical folding of Dynkin diagrams and reduction of
affine Toda theories}}

\vskip 1.5cm

{\large {\bf S. Pratik  Khastgir
\footnote{Monbusho Fellow.}
and Ryu Sasaki
\footnote{Supported partially by the grant-in-aid for Scientific
Research,
Priority Area 231 ``Infinite Analysis'' and General Research (C) in
Physics, Japan Ministry of Education.}}}

\vskip 0.5cm

{\sl Yukawa Institute for Theoretical Physics,} \\
{\sl Kyoto University,} \\
{\sl Kyoto \ 606-01, Japan.}

\end{center}

\vspace{1 cm}

\begin{abstract}
The equation of motion of affine Toda field theory is a coupled equation
for $r$ fields,
 $r$ is the rank of the underlying Lie algebra.
Most of the theories admit reduction,
in which the equation is satisfied by fewer than $r$ fields.
The reductions in the existing literature are achieved by identifying
(folding) the points in the Dynkin diagrams which are connected by
symmetry (automorphism). In this paper we present many new reductions.
In other words the symmetry of affine Dynkin diagrams could be extended
and it leads to non-canonical foldings. We investigate these reductions
in detail and formulate general rules for possible reductions. We will show
that eventually most of the theories end up in $a_{2n}^{(2)}$ that is
the theory cannot have a further dimension $m$ reduction where $m<n$.

\end{abstract}

\vspace{1 cm}
\newcommand{\NN}{\nonumber}
\renewcommand{\theequation}{\arabic{section}.\arabic{equation}}
\newcommand\p{^\prime}
\newcommand\pp{^{\prime\prime}}
\def\ts{\thinspace}
\newcommand\wt{\widetilde}
\newcommand\wh{\widehat}
\newcommand\ccon{^*}
\newcommand\beq{\begin{equation}}
\newcommand\eeq{\end{equation}}
\newcommand{\rref}[1]{(\ref{#1})}

\newcommand\bear{\begin{array}}
\newcommand\enar{\end{array}}
\newcommand\Bear{\begin{eqnarray}}
\newcommand\Enar{\end{eqnarray}}
\newcommand\Bears{\begin{eqnarray*}}
\newcommand\Enars{\end{eqnarray*}}
\newcommand\lang{\langle}
\newcommand\llang{\langle\!\langle}
\newcommand\rang{\rangle}
\newcommand\rrang{\rangle\!\rangle}
\newcommand\bas{\backslash}
\newcommand\Sol{\lq\lq soliton"\ }
\newcommand\Sols{\lq\lq solitons"\ }
\newcommand\aSol{\lq\lq anti-soliton"\ }
\newcommand\Do{\lq\lq dimension one"\ }
\newcommand\Br{\lq\lq breather"\ }
\newcommand\Brs{\lq\lq breathers"\ }
\newcommand\Bo{\lq\lq bootstrap"\ }
\newcommand\El{\lq\lq elementary"\ }
\newcommand\Bou{\lq\lq bound state"\ }
\newcommand\Bous{\lq\lq bound states"\ }
\newcommand\Sp{\lq\lq spinor"\ }
\newcommand\Sps{\lq\lq spinors"\ }
\def\hd#1{( #1 )\hat{}}
\def\hcd#1{\{ #1 \}\hat{}}
\def\bd#1{( #1 )\breve{}}
\def\bcd#1{\{ #1 \}\breve{}}
\def\tB{\tilde B}
\def\td#1{( #1 )\tilde{}}
\def\ttd#1{\{ #1 \}\tilde{}}
\def\unit#1{\langle\ #1\ \rangle}
\newcommand\Zam{Zamolodchikov}
\newcommand \ZZ {A. B. \Zam\  and Al. B. \Zam, {\it Ann. Phys.}
{\bf 120} (1979)  253}
\newcommand\sh{{\rm sh}}
\newcommand\ch{{\rm ch}}
\newcommand\algg{{\bf g }}
\newcommand\atnt{$a_{2n}^{(2)}$}
\newcommand\aont{$a_{2n-1}^{(2)}$}
\newcommand\dfth{$d_{4}^{(3)}$}
\newcommand\ano{$a_{n}^{(1)}$}
\newcommand\bno{$b_{n}^{(1)}$}
\newcommand\cno{$c_{n}^{(1)}$}
\newcommand\dnt{$d_{n+1}^{(2)}$}
\newcommand\dno{$d_{n}^{(1)}$}
\newcommand\esio{$e_{6}^{(1)}$}
\newcommand\eseo{$e_{7}^{(1)}$}
\newcommand\eeho{$e_{8}^{(1)}$}
\newcommand\esit{$e_{6}^{(2)}$}
\newcommand\ffo{$f_{4}^{(1)}$}
\newcommand\gto{$g_{2}^{(1)}$}
\newcommand\alp{\alpha_}
\newcommand\bet{\beta_}
\let \rf=\refmark
\def\NPrefmark#1{\attach{\scriptstyle [ #1]  }}

\newcommand\ASPM{{\it Advanced Studies in Pure Mathematics\ts}}
\newcommand\CMP{{\it Comm.\ts Math.\ts Phys.\ts}}
\newcommand\CPC{{\it Computer\ts Phys.\ts Comm.\ts}}
\newcommand\FAP{{\it Funct.\ts Analy.\ts Appl.\ts}}
\newcommand\IJMP{{\it Int.\ts J.\ts Mod.\ts Phys.\ts}}
\newcommand\INVM{{\it Inv.\ts Math.\ts}}
\newcommand\JMP{{\it J.\ts Math.\ts Phys.\ts}}
\newcommand\JP{{\it J.\ts Phys.\ts}}
\newcommand\MPL{{\it Mod.\ts Phys.\ts Lett.\ts}}
\newcommand\NP{{\it Nucl.\ts Phys.\ts}}
\newcommand\PL{{\it Phys.\ts Lett.\ts}}
\newcommand\PJM{{\it Pacific\ts J.\ts Math.\ts}}
\newcommand\PNAS{{\it Proc.\ts Natl.\ts Acad.\ts Sci.\ts USA\ts}}
\newcommand\PR{{\it Phys.\ts Rev.\ts}}
\newcommand\PRL{{\it Phys.\ts Rev.\ts Lett.\ts}}
\newcommand\PSPM{{\it Proc.\ts Symp.\ts Pure\ts Math.\ts}}
\newcommand\PTP{{\it Prog.\ts Theor.\ts Phys.\ts}}
\newcommand\TAMS{{\it Trans.\ts Amer.\ts Math.\ts Soc.\ts}}
\newcommand\TMP{{\it Theor.\ts Math.\ts Phys.\ts}}
\newcommand\SJNP{{\it Sov.\ts J.\ts Nucl.\ts Phys.\ts}}
\newcommand\Zm{Zamolodchikov}
\newcommand\AZm{A.\ts B.\ts \Zm}
\newcommand\dur{H.\ts W.\ts Braden, E.\ts Corrigan, P.\ts E.\ts Dorey
and R.\ts Sasaki}


\font\dynkfont=cmsy10 scaled\magstep4    \skewchar\dynkfont='60
\def\dynk{\textfont2=\dynkfont}
\def\hr#1,#2;{\dimen0=.4pt\advance\dimen0by-#2pt
              \vrule width#1pt height#2pt depth\dimen0}
\def\vr#1,#2;{\vrule height#1pt depth#2pt}
\def\blb#1#2#3#4#5
            {\hbox{\ifnum#2=0\hskip11.5pt
           \else\ifnum#2=1\hr13,5.4;\hskip-1.5pt
           \else\ifnum#2=2\hr13.5,7.6;\hskip-13.5pt
                        \hr13.5,3.2;\hskip-2pt
                   \else\ifnum#2=3\hr13.7,8.4;\hskip-13.7pt
                                  \hr13,5.4;\hskip-13pt
                                  \hr13.7,2.4;\hskip-2.2pt
                   \else\ifnum#2=7\hr13.7,8.4;\hskip-13.7pt
                                  \hr13.2,6.4;\hskip-13.2pt
                                  \hr13.2,4.4;\hskip-13.2pt
                                  \hr13.7,2.4;\hskip-2.2pt
                   \else\ifnum#2=8\hr13.7,8.4;\hskip-13.7pt
                                  \hr13.2,6.4;\hskip-13.2pt
                                  \hr13.2,4.4;\hskip-13.2pt
                 \hr13.7,2.4;\hskip-8.2pt$\rangle$\hskip-2.0pt
                   \else\ifnum#2=4\hr13.5,7.6;\hskip-13.5pt
                \hr13.5,3.2;\hskip-8pt$\rangle$\hskip-1.8pt
                   \else\ifnum#2=5\hr30,5.4;\hskip-1.5pt
                   \else\ifnum#2=6\hr13.7,8.4;\hskip-13.7pt
                                  \hr13,5.4;\hskip-13pt
               \hr13.7,2.4;\hskip-8.2pt$\rangle$\hskip-2.0pt
                      \fi\fi\fi\fi\fi\fi\fi\fi\fi
                   $#1$
                   \ifnum#4=0
                   \else\ifnum#4=1\hskip-9.2pt\vr22,-9;\hskip8.8pt
                   \else\ifnum#4=2\hskip-10.9pt\vr22,-8.75;\hskip3pt
                                  \vr22,-8.75;\hskip7.1pt
                   \else\ifnum#4=3\hskip-12.6pt\vr22,-8.5;\hskip3pt
                                  \vr22,-9;\hskip3pt
                                  \vr22,-8.5;\hskip5.4pt
                   \else\ifnum#4=5\hskip-9.2pt\vr39,-9;\hskip8.8pt
                                  \fi\fi\fi\fi\fi
                   \ifnum#5=0
                   \else\ifnum#5=1\hskip-9.2pt\vr1,12;\hskip8.8pt
                   \else\ifnum#5=2\hskip-10.9pt\vr1.25,12;\hskip3pt
                                  \vr1.25,12;\hskip7.1pt
                   \else\ifnum#5=3\hskip-12.6pt\vr1.5,12;\hskip3pt
                                  \vr1,12;\hskip3pt
                                  \vr1.5,12;\hskip5.4pt
                   \else\ifnum#5=5\hskip-9.2pt\vr1,29;\hskip8.8pt
                                   \fi\fi\fi\fi\fi

                   \ifnum#3=0\hskip8pt
                   \else\ifnum#3=1\hskip-5pt\hr13,5.4;
                   \else\ifnum#3=2\hskip-5.5pt\hr13.5,7.6;
                                  \hskip-13.5pt\hr13.5,3.2;
                   \else\ifnum#3=3\hskip-5.7pt\hr13.7,8.4;
                                  \hskip-13pt\hr13,5.4;
                                  \hskip-13.7pt\hr13.7,2.4;
                  \else\ifnum#3=7\hskip-5.7pt\hr13.7,8.4;
                                 \hskip-13.2pt\hr13.2,6.4;
                                 \hskip-13.2pt\hr13.2,4.4;
                                  \hskip-13.7pt\hr13.7,2.4;
                  \else\ifnum#3=8\hskip-5.7pt\hr13.7,8.4;
                                 \hskip-13.2pt\hr13.2,6.4;
                                 \hskip-13.2pt\hr13.2,4.4;
                    \hskip-13.9pt$\langle$\hskip-8pt\hr13.7,2.4;
                   \else\ifnum#3=4\hskip-5.5pt\hr13.5,7.6;
                                  \hskip-13.5pt$\langle$\hskip-8.2pt
                                  \hr13.5,3.2;
                  \else\ifnum#3=5\hskip-5pt\hr30,5.4;
                   \else\ifnum#3=6\hskip-5.7pt\hr13.7,8.4;
                                  \hskip-13pt\hr13,5.4;
                       \hskip-13.9pt$\langle$\hskip-8.0pt\hr13.7,2.4;
                                 \fi\fi\fi\fi\fi\fi\fi\fi\fi
                   }}
\def\blob#1#2#3#4#5#6#7{\hbox
{$\displaystyle\mathop{\blb#1#2#3#4#5 }_{#6}\sp{#7}$}}
\def\up#1#2{\dimen1=33pt\multiply\dimen1by#1
                  \hbox{\raise\dimen1\rlap{#2}}}
\def\uph#1#2{\dimen1=17.5pt\multiply\dimen1by#1
                  \hbox{\raise\dimen1\rlap{#2}}}
\def\dn#1#2{\dimen1=33pt\multiply\dimen1by#1
                   \hbox{\lower\dimen1\rlap{#2}}}
\def\dnh#1#2{\dimen1=17.5pt\multiply\dimen1by#1
                    \hbox{\lower\dimen1\rlap{#2}}}

\def\rlbl#1{\kern-8pt\raise3pt\hbox{$\scriptstyle #1$}}
\def\llbl#1{\raise3pt\llap{\hbox{$\scriptstyle #1$\kern-8pt}}}
\def\elbl#1{\kern3pt\lower4.5pt\hbox{$\scriptstyle #1$}}
\def\lelbl#1{\rlap{\hbox{\kern-9pt\raise2.5pt\hbox{{$\scriptstyle #1$}}}}}

\def\wht#1#2#3#4{\blob\circ#1#2#3#4{}{}}
\def\blk#1#2#3#4{\blob\bullet#1#2#3#4{}{}}
\def\whtd#1#2#3#4#5{\blob\circ#1#2#3#4{#5}{}}
\def\blkd#1#2#3#4#5{\blob\bullet#1#2#3#4{#5}{}}
\def\whtu#1#2#3#4#5{\blob\circ#1#2#3#4{}{#5}}
\def\blku#1#2#3#4#5{\blob\bullet#1#2#3#4{}{#5}}
\def\whtr#1#2#3#4#5{\blob\circ#1#2#3#4{}{}\rlbl{#5}}
\def\whtre#1#2#3#4#5{\blob\circ#1#2#3#4{}{}\rlbl{#5}}
\def\blkr#1#2#3#4#5{\blob\bullet#1#2#3#4{}{}\rlbl{#5}}
\def\whtl#1#2#3#4#5{\llbl{#5}\blob\circ#1#2#3#4{}{}}
\def\whtle#1#2#3#4#5{\llbl{#5}\blob\circ#1#2#3#4{}{}}
\def\blkl#1#2#3#4#5{\llbl{#5}\blob\bullet#1#2#3#4{}{}}

\def\rwng{\hbox{$\vbox{\offinterlineskip{
  \hbox{\phantom{}\kern6pt{$\circ$}}\kern-2.5pt\hbox{$\Biggr/$}\kern-0.5pt
  \hbox{\phantom{}\kern-5pt$\circ$}\kern-3.0pt\hbox{$\Biggr\backslash$}
  \kern-1.5pt\hbox{\phantom{}\kern6pt{$\circ$}} }}$}}

\def\lwng{\hbox{$\vbox{\offinterlineskip{ \hbox{$\circ$}
  \kern-3.0pt\hbox{\phantom{}\kern6.0pt{$\Biggr\backslash$}}
  \kern-0.5pt\hbox{\phantom{}\kern11pt{$\circ$}}\kern-3.5pt
  \hbox{\phantom{}\kern5.0pt {$\Biggr/$}}\kern-1.0pt\hbox{$\circ$} }}$}}

\def\drwng#1#2#3{\hbox{$\vcenter{ \offinterlineskip{
  \hbox{\phantom{}\kern6pt{$\circ^{\elbl{#3}}$}}
  \kern-2.5pt\hbox{$\Biggr/$}\kern-0.5pt
  \hbox{\phantom{}\kern-5pt$\circ^{ \elbl{#1}}$}
  \kern-3.0pt\hbox{$\Biggr\backslash$}
  \kern-1.5pt\hbox{\phantom{}\kern6pt{$\circ^{\elbl{#2}}$}}  } }$}}

\def\dlwng#1#2#3{\hbox{$\vcenter{\offinterlineskip{ \hbox{$\lelbl{#1}\circ$}
  \kern-3.0pt\hbox{\phantom{}\kern6.0pt{$\Biggr\backslash$}}
  \kern-0.5pt\hbox{\phantom{}\kern11pt{$\lelbl{#2}\circ$}}\kern-3.5pt
  \hbox{\phantom{}\kern5.0pt {$\Biggr/$}}\kern-1.0pt%
   \hbox{$\lelbl{#3}\circ$}}}$}
}

\def\rde#1#2#3{\raisebox{.5pt}{\hbox{\phantom{}\kern-4pt\hbox{$\vcenter
{\offinterlineskip\hbox{
               \raise 4.5pt\hbox{\vrule height0.4pt width13pt depth0pt}
                \kern-1pt\vbox{ \hbox{\drwng{#1}{#2}{#3}}} }}$  }} }}

\def\lde#1#2#3{\raisebox{.5pt}{\hbox{$\vcenter{\offinterlineskip  \hbox{
         \dlwng{#1}{#2}{#3}\kern-5.2pt\lower0.4pt\hbox{$\vcenter{\hrule
 width13pt}$}
               \kern-8pt\phantom{}   }}  $}}}

\def\ldet#1#2#3{\hbox{$\vcenter{\offinterlineskip  \hbox{
               \dlwng{#1}{#2}{#3}\kern14pt\lower0.4pt\hbox{$\vcenter{
          \hskip-20pt\hr13.5,5.7;\hskip-13.5pt\hr13.5,1.3;}$}
               \kern-25.8pt\phantom{}   }}  $}}

\def\dre{\rde{}{}{}}

\def\dle{\lde{}{}{}}

\def\rwngb{\hbox{$\vbox{\offinterlineskip{
\hbox{\phantom{}\kern6pt{$\bullet$}}\kern-2.5pt\hbox{$\Biggr/$}\kern-0.5pt
  \hbox{\phantom{}\kern-5pt$\bullet$}\kern-3.0pt\hbox{$\Biggr\backslash$}
  \kern-1.5pt\hbox{\phantom{}\kern6pt{$\bullet$}} }}$}}

\def\lwngb{\hbox{$\vbox{\offinterlineskip{ \hbox{$\bullet$}
  \kern-3.0pt\hbox{\phantom{}\kern6.0pt{$\Biggr\backslash$}}
  \kern-0.5pt\hbox{\phantom{}\kern11pt{$\bullet$}}\kern-3.5pt
  \hbox{\phantom{}\kern5.0pt {$\Biggr/$}}\kern-1.0pt\hbox{$\bullet$} }}$}}

\def\dbrwng#1#2#3{\hbox{$\vcenter{ \offinterlineskip{
  \hbox{\phantom{}\kern6pt{$\bullet^{\elbl{#3}}$}}
  \kern-2.5pt\hbox{$\Biggr/$}\kern-0.5pt
  \hbox{\phantom{}\kern-5pt$\bullet^{ \elbl{#1}}$}
  \kern-3.0pt\hbox{$\Biggr\backslash$}
  \kern-1.5pt\hbox{\phantom{}\kern6pt{$\bullet^{\elbl{#2}}$}}  } }$}}

\def\dblwng#1#2#3{\hbox{$\vcenter{\offinterlineskip{
   \hbox{$\lelbl{#1}\bullet$}
  \kern-3.0pt\hbox{\phantom{}\kern6.0pt{$\Biggr\backslash$}}
  \kern-0.5pt\hbox{\phantom{}\kern11pt{$\lelbl{#2}\bullet$}}\kern-3.5pt
  \hbox{\phantom{}\kern5.0pt
 {$\Biggr/$}}\kern-1.0pt\hbox{$\lelbl{#3}\bullet$}}}$} }

\def\rbde#1#2#3{\hbox{\phantom{}\kern-4pt\hbox{$\vcenter{\offinterlineskip
 \hbox{
               \raise 4.5pt\hbox{\vrule height0.4pt width13pt depth0pt}
                \kern-1pt\vbox{ \hbox{\dbrwng{#1}{#2}{#3}}} }}$  }}  }

\def\lbde#1#2#3{\hbox{$\vcenter{\offinterlineskip  \hbox{
        \dblwng{#1}{#2}{#3}\kern-4.2pt\lower0.4pt\hbox{$\vcenter{\hrule
 width13pt}$}
               \kern-8pt\phantom{}   }}  $}}

\def\dbre{\rbde{}{}{}}

\def\dble{\lbde{}{}{}}

\def\p#1{\phantom{-}#1}
\def\ins{\phantom{\ldots}}

\def\eddgiu#1.#2.#3.{\dynk \whtu0100{#1}\whtu1300{#2}\whtu6000{#3}}
\def\eddgid#1.#2.#3.{\dynk \whtd0100{#1}\whtd1300{#2}\whtd6000{#3}}
\def\eddgiid#1.#2.#3.{\dynk  \whtd0300{#1}\whtd6100{#2}\whtd1000{#3}}

\def\eddfiu#1.#2.#3.#4.#5.{\dynk
 \whtu0100{#1}\whtu1100{#2}\whtu1200{#3}\whtu4100{#4}\whtu1000{#5}}

\def\eddfid#1.#2.#3.#4.#5.{\dynk
 \whtd0100{#1}\whtd1100{#2}\whtd1200{#3}\whtd4100{#4}\whtd1000{#5}}

\def\eddfiiu#1.#2.#3.#4.#5.{\dynk
 \whtu0100{#1}\whtu1200{#2}\whtu4100{#3}\whtu1100{#4}\whtu1000{#5}}
\def\eddfiid#1.#2.#3.#4.#5.{\dynk
 \whtu0100{#1}\whtd1200{#2}\whtd4100{#3}\whtu1100{#4}\whtd1000{#5}}

\def\ddanu#1.#2.#3.#4.#5.{\dynk \whtu0100{#1}\whtu1100{#2}\whtu1100{#3}%
                          \cdots\whtu1100{#4}\whtu1000{#5}}
\def\ddanuf#1.#2.#3.#4.{\dynk \whtd0100{#1}\whtu1100{#2}\cdots%
                           \whtu1100{#3}\whtu1000{#4}}
\def\ddanuuf#1.#2.#3.#4.{\dynk \whtu0100{#1}\whtu1100{#2}\cdots%
                           \whtu1100{#3}\whtu1000{#4}}
\def\ddanufd#1.#2.#3.#4.{\dynk \whtd0100{#1}\whtu1100{#2}\cdots%
                           \whtu1100{#3}\whtr1005{#4}}
\def\ddandf#1.#2.#3.#4.{\dynk \whtu0100{#1}\whtd1100{#2}\cdots%
                           \whtd1100{#3}\whtd1000{#4}}
\def\ddanddf#1.#2.#3.#4.{\dynk \whtd0100{#1}\whtd1100{#2}\cdots%
                           \whtd1100{#3}\whtd1000{#4}}
\def\ddandfu#1.#2.#3.#4.{\dynk \whtu0100{#1}\whtd1100{#2}\cdots%
                           \whtd1100{#3}\whtr1050{#4}}
\def\ddcnds#1.#2.#3.#4.#5.#6{\dynk \whtd0200{#1}\whtd4100{#2}%
                          \whtd1100{#3}\cdots%
                           \whtd1100{#4}\whtd1400{#5}\whtd2000{#6}}

\def\eddanu#1.#2.#3.#4.#5.{\dynk \whtu0100{#1}\whtu1100{#2}%
               \up1{\whtr0000{#3}}\cdots\whtu1100{#4}\whtu1000{#5}}
\def\eddand#1.#2.#3.#4.#5.{\dynk \whtd0100{#1}\whtd1100{#2}%
                \up1{\whtr0000{#3}}\cdots\whtd1100{#4}\whtd1000{#5}}
\def\ddand#1.#2.#3.#4.#5.{\dynk \whtd0100{#1}\whtd1100{#2}\whtd1100{#3}%
               \cdots\whtd1100{#4}\whtd1000{#5}}
\def\andfive#1.#2.#3.#4.#5.{\dynk \whtu0100{#1}\whtu1100{#2}\whtu1100{#3}%
                           \whtu1100{#4}\whtu1000{#5}}
\def\andthr#1.#2.#3.{\dynk \whtu0100{#1}\whtu1100{#2}\whtu1000{#3}}

\def\eddanid#1.#2.#3.#4.#5.{\dynk \whtd0200{#1}\whtd4100{#2}%
                           \whtd1100{#3}\cdots\whtd1200{#4}\whtd4000{#5}}
\def\eddanidr#1.#2.#3.#4.#5.{\dynk \whtd0200{#1}\whtd4100{#2}%
                           \cdots\whtd1100{#3}\whtd1200{#4}\whtd4000{#5}}

\def\eddaniid#1.#2.#3.#4.#5.#6.{\hbox{$\vcenter{\hbox
         {\dynk\hbox{$ \lde{#1}{#2}{#3}\whtd1100{#4}\cdots%
          \whtd1400{#5}\whtd2000{#6} $}} }$}}

\def\eddaiii#1.#2.#3.{\dynk\whtd0400{#1}\whtd2200{#2}\whtd4000{#3}}
\def\eddaiiif#1.#2.#3.#4.{\dynk\whtd0400{#1}\whtd2100{#2}%
                    \whtd1200{#3}\whtd4000{#4}}
\def\eddciii#1.#2.#3.{\dynk\whtd0200{#1}\whtd4400{#2}\whtd2000{#3}}

\def\eddbnd#1.#2.#3.#4.#5.#6.{\dynk \lde{#1}{#2}{#3}\whtd1100{#4}\cdots%
                           \whtd1200{#5}\whtd4000{#6}}
\def\eddbndt#1.#2.#3.#4.{\dynk \ldet{#1}{#2}{#3}\hskip-444pt\whtr4000{#4}}

\def\ncddlr#1.#2.#3.#4.#5.{\dynk \whtu0101{#1}\whtu1100{#2}\whtu1100{#3}%
                           \cdots\whtu1100{#4}\whtu1001{#5}}
\def\ncdulr#1.#2.#3.#4.#5.{\dynk \whtd0110{#1}\whtd1100{#2}\whtd1100{#3}%
                           \cdots\whtd1100{#4}\whtd1010{#5}}
\def\ncddrd#1.#2.#3.#4.#5.{\dynk \whtd0100{#1}\whtu1100{#2}\whtu1100{#3}%
                           \cdots\whtu1100{#4}\whtu1005{#5}}
\def\ncddru#1.#2.#3.#4.#5.{\dynk \whtu0100{#1}\whtd1100{#2}\whtd1100{#3}%
                           \cdots\whtd1100{#4}\whtd1050{#5}}
\def\ncddrdu#1.#2.#3.#4.#5.{\dynk \whtu0100{#1}\whtu1100{#2}\whtu1100{#3}%
                         \cdots\whtu1100{#4}\whtu1005{#5}}
\def\ncddrud#1.#2.#3.#4.#5.{\dynk \whtd0100{#1}\whtd1100{#2}\whtd1100{#3}%
                            \cdots\whtd1100{#4}\whtd1050{#5}}
\def\ncddld#1.#2.#3.#4.#5.{\dynk \whtl0105{#1}\whtu1100{#2}\whtu1100{#3}%
                          \cdots\whtu1100{#4}\whtd1000{#5}}
\def\ncddlu#1.#2.#3.#4.#5.{\dynk \whtl0150{#1}\whtd1100{#2}\whtd1100{#3}%
                           \cdots\whtd1100{#4}\whtu1000{#5}}
\def\ncanur#1.#2.#3.#4.#5.{\dynk \whtu0101{#1}\whtu1100{#2}\whtu1100{#3}%
                           \cdots\whtu1100{#4}\whtd1010{#5}}
\def\ncandr#1.#2.#3.#4.#5.{\dynk \whtd0110{#1}\whtd1100{#2}\whtd1100{#3}%
                           \cdots\whtd1100{#4}\whtu1001{#5}}

\def\eddcnd#1.#2.#3.#4.#5.{\dynk \whtd0200{#1}\whtd4100{#2}\whtd1100{#3}
       \cdots \whtd1400{#4}\whtd2000{#5}}

\def\dddnu#1.#2.#3.#4.#5.#6.{\hbox{$\vcenter{\hbox
         {\dynk\hbox{$ \whtu0100{#1}\whtu1100{#2}\cdots%
          \whtu1100{#3}\rde{#4}{#5}{#6} $}}  }$}}
\def\dddnd#1.#2.#3.#4.#5.#6.{\hbox{$\vcenter{\hbox
         {\dynk\hbox{$ \whtd0100{#1}\whtd1100{#2}\cdots%
          \whtd1100{#3}\rde{#4}{#5}{#6} $}} }$}}
\def\dddiv#1.#2.#3.#4.{\hbox{$\vcenter{\hbox
         {\dynk\hbox{$ \whtu0100{#1}\rde{#2}{#3}{#4}
              $}}  }$}}
\def\edddiv#1.#2.#3.#4.{\hbox{$\vcenter{\hbox{\dynk\hbox{$\whtl0100{#1}
\up1{\whtl0001{#2}}\dn1{\whtl0010{#4}}\wht1111\whtr1000{#3} $}}}$}}

\def\edddnu#1.#2.#3.#4.#5.#6.#7.#8.{\hbox{$\vcenter{\hbox
         {\dynk\hbox{$ \lde{#1}{#2}{#3}\whtu1100{#4}\cdots%
          \whtu1100{#5}\rde{#6}{#7}{#8} $}}  }$}}
\def\edddnd#1.#2.#3.#4.#5.#6.#7.#8.{\hbox{$\vcenter{\hbox
         {\dynk\hbox{$ \lde{#1}{#2}{#3}\whtd1100{#4}\cdots%
          \whtd1100{#5}\rde{#6}{#7}{#8} $}} }$}}
\def\edddndf#1.#2.#3.#4.#5.#6.{\hbox{$\vcenter{\hbox
         {\dynk\hbox{$ \lde{#1}{#2}{#3}\rde{#4}{#5}{#6} $}} }$}}

\def\edddnds#1.#2.#3.#4.#5.#6.#7.#8.#9.{\hbox{$\vcenter{\hbox
{\dynk\hbox{$ \lde{#1}{#2}{#3}\whtd1100{#4}\cdot\cdot\whtd1100{#5}\cdot%
      \cdot\whtd1100{#6}\rde{#7}{#8}{#9} $}} }$}}
\def\eddanod#1.#2.#3.#4.#5.#6.{\hbox{$\vcenter{\hbox
         {\dynk\hbox{$ \whtd0200{#1}\whtd4100{#2}\cdots%
          \whtd1100{#3}\rde{#4}{#5}{#6} $}} }$}}

\def\edddniid#1.#2.#3.#4.#5.{\hbox{$\vcenter{\hbox
         {\dynk\hbox{$ \whtd0400{#1}\whtd2100{#2}\whtd1100{#3}\cdots%
          \whtd1200{#4}\whtd4000{#5} $}} }$}}

\def\edddniiu#1.#2.#3.#4.#5.{\hbox{$\vcenter{\hbox
         {\dynk\hbox{$ \blku0200{#1}\whtu2100{#2}\whtu1100{#3}\cdots%
          \whtu1200{#4}\blku2000{#5} $}} }$}}

\def\ddei#1.#2.#3.#4.#5.#6.{\hbox{$\vcenter{\hbox
       {\dynk \whtd0100{#1}\whtd1100{#3}%
       \up1{\whtr0001{#2}}\whtd1110{#4}\whtd1100{#5}\whtd1000{#6}} }$}}

\def\eddei#1.#2.#3.#4.#5.#6.#7.{\hbox{$\vcenter{\hbox
       {\dynk \whtu0100{#1}\whtu1100{#3}%
       \up1{\whtr0011{#2}}\up2{\whtr0001{#7}}\whtd1110{#4}\whtu1100{#5}%
       \whtu1000{#6}} }$}}

\def\ncdddt#1.#2.{\dynk\whtu0400{#1}\whtu2001{#2}}
\def\ncandrt#1.#2.{\dynk\whtd0110{#1}\whtu1001{#2}}
\def\ncanurt#1.#2.{\dynk\whtu0101{#1}\whtd1010{#2}}
\def\ncddet#1.#2.{\dynk\whtu0400{#1}\whtu2000{#2}}
\def\ncddut#1.#2.{\dynk\whtd0400{#1}\whtd2010{#2}}
\def\ncdduot#1.#2.{\dynk\whtd0210{#1}\whtd4000{#2}}
\def\ncddct#1.#2.{\hbox{\dynk\whtu0200{\rotatebox{45}{$\scriptstyle#1$}}%
                    \whtu4000{\rotatebox{45}{$\scriptstyle#2$}}}}
\def\ncddcot#1.#2.{\dynk\whtd0400{\rotatebox{45}{$\scriptstyle#1$}}%
                    \whtd2000{\rotatebox{45}{$\scriptstyle#2$}}}
\def\ncddcst#1.#2.{\dynk\whtu0400{\rotatebox{135}{$\scriptstyle#1$}}%
                    \whtu2000{\rotatebox{135}{$\scriptstyle#2$}}}

\def\rronit#1.{\rotatebox{315}
       {\dynk\whtr5005{\rotatebox{45}{$\scriptstyle #1$}}}}
\def\laronit#1.#2.{\rotatebox{315}{\ncddct#1.#2.}}
\def\raronit#1.#2.{\rotatebox{315}{\ncddcot#1.#2.}}
\def\rarsnit#1.#2.{\rotatebox{225}{\ncddcst#1.#2.}}

\def\ncddd#1.#2.#3.#4.#5.{\dynk\whtu0400{#1}\whtu2100{#2}\whtu1100{#3}%
           \cdots\whtu1100{#4}\whtu1001{#5}}
\def\ncdde#1.#2.#3.#4.#5.{\dynk\whtu0400{#1}\whtu2100{#2}\whtu1100{#3}%
           \cdots\whtu1100{#4}\whtu1000{#5}}
\def\ncdded#1.#2.#3.#4.#5.{\dynk\whtl0400{#1}\whtd2100{#2}\whtd1100{#3}%
           \cdots\whtd1100{#4}\whtd1000{#5}}
\def\ncddeo#1.#2.#3.#4.#5.{\dynk\whtd0100{#1}\whtd1100{#2}\whtd1100{#3}%
           \cdots\whtd1200{#4}\whtd4000{#5}}
\def\ncddeof#1.#2.#3.#4.{\dynk\whtd0100{#1}\whtd1100{#2}%
           \cdots\whtd1200{#3}\whtr4000{#4}}
\def\ncddu#1.#2.#3.#4.#5.{\dynk\whtd0400{#1}\whtd2100{#2}\whtd1100{#3}%
           \cdots\whtd1100{#4}\whtd1010{#5}}
\def\ncdduo#1.#2.#3.#4.#5.{\dynk\whtd0110{#1}\whtd1100{#2}\whtd1100{#3}%
           \cdots\whtd1200{#4}\whtd4000{#5}}
\def\ncddc#1.#2.#3.#4.#5.{\dynk\whtu0200{#1}\whtu4100{#2}\whtu1100{#3}%
           \cdots\whtu1100{#4}\whtu1000{#5}}
\def\ncdddc#1.#2.#3.#4.#5.{\dynk\whtd0200{#1}\whtd4100{#2}\whtd1100{#3}%
           \cdots\whtd1100{#4}\whtd1000{#5}}
\def\ncddcu#1.#2.#3.#4.#5.{\dynk\whtl0200{#1}\whtd4100{#2}\whtd1100{#3}%
           \cdots\whtd1100{#4}\whtr1050{#5}}
\def\ncddcd#1.#2.#3.#4.#5.{\dynk\whtu0200{#1}\whtu4100{#2}\whtu1100{#3}%
           \cdots\whtu1100{#4}\whtu1005{#5}}
\def\ncddco#1.#2.#3.#4.#5.{\dynk\whtd0100{#1}\whtd1100{#2}\cdots%
            \whtd1100{#3}\whtd1400{#4}\whtd2000{#5}}
\def\ncdfr#1.#2.#3.#4.#5.#6.{\ncddc#1.#2.#3.#4.#5.
           \hskip-42.5pt\dynk\rotatebox{315}
          {\whtu5005{\rotatebox{45}{$\scriptstyle #6$}}}}
\def\ncdfrd#1.#2.#3.#4.#5.#6.{\ncdde#1.#2.#3.#4.#5.
           \hskip-42.5pt\dynk\rotatebox{315}
          {\whtr5005{\rotatebox{45}{$\scriptstyle #6$}}}}
\def\ncdfrdc#1.#2.#3.#4.#5.#6.{\ncddc#1.#2.#3.#4.#5.
           \hskip-42.5pt\dynk\rotatebox{315}
          {\whtr5005{\rotatebox{45}{$\scriptstyle #6$}}}}
\def\ncdfrdl#1.#2.#3.#4.#5.#6.{\ncddlu#1.#2.#3.#4.#5.
           \hskip-42.5pt\dynk\rotatebox{315}
          {\whtr5005{\rotatebox{45}{$\scriptstyle #6$}}}}

\def\ncdfl#1.#2.#3.#4.#5.#6.{\hbox{\lronit#1.\hskip-39.5pt
                   \raisebox{13pt}{$\ddand#2.#3.#4.#5.#6.$}}}
\def\ncdfal#1.#2.#3.#4.#5.{\hbox{\lronit#1.\hskip-39.5pt
                   \raisebox{13pt}{$\ddandf#2.#3.#4.#5.$}}}
\def\ncdfalu#1.#2.#3.#4.#5.{\hbox{\lronit#1.\hskip-39.5pt
                   \raisebox{13pt}{$\ddandfu#2.#3.#4.#5.$}}}

\def\ncdfar#1.#2.#3.#4.#5.{\ddanuf#1.#2.#3.#4.
  \hskip-42.5pt\dynk\rotatebox{315}{\whtr5005{\rotatebox{45}
           {$\scriptstyle #5$}}}}
\def\ncdfaur#1.#2.#3.#4.#5.{\ddanuuf#1.#2.#3.#4.
  \hskip-42.5pt\dynk\rotatebox{315}{\whtu5005{\rotatebox{45}
           {$\scriptstyle #5$}}}}

\def\lronit#1.{\rotatebox{315}
       {\dynk\whtl0550{\rotatebox{45}{$\scriptstyle #1$}}}}
\def\daone#1.#2.{\dynk\whtd0400{#1}\whtd4000{#2}}

\def\datwot#1.#2.{\dynk\whtu0700{#1}\whtu8000{#2}}

\def\datwon#1.#2.#3.#4.#5.#6.{\dynk \whtd0200{#1}\whtd4100{#2}%
         \whtd1100{#3}\whtd1100{#4} \cdots\whtd1200{#5}\whtd4000{#6}}
\def\datwono#1.#2.#3.#4.#5.#6.{\dynk \whtd0400{#1}\whtd2100{#2}%
        \whtd1100{#3}\cdots \whtd1100{#4}\whtd1400{#5}\whtd2000{#6}}
\def\datwonl#1.#2.#3.#4.#5.#6.{\dynk \whtd0200{#1}\whtd4100{#2}%
         \whtd1100{#3}\cdots \whtd1100{#4} \whtd1200{#5}\whtd4000{#6}}

\def\ddeii#1.#2.#3.#4.#5.#6.#7.{\hbox{$\vcenter{\hbox
       {\dynk \whtd0100{#1}\whtd1100{#3}%
       \up1{\whtr0001{#2}}\whtd1110{#4}\whtd1100{#5}\whtd1100{#6}%
       \whtd1000{#7}} }$}}

\def\eddeii#1.#2.#3.#4.#5.#6.#7.#8.{\hbox{$\vcenter{\hbox
       {\dynk \whtu0100{#8}\whtu1100{#1}\whtu1100{#3}%
       \up1{\whtr0001{#2}}\whtd1110{#4}\whtu1100{#5}\whtu1100{#6}%
       \whtu1000{#7}} }$}}

\def\ddeiii#1.#2.#3.#4.#5.#6.#7.#8.{\hbox{$\vcenter{\hbox
       {\dynk \whtd0100{#1}\whtd1100{#3}%
       \up1{\whtr0001{#2}}\whtd1110{#4}\whtd1100{#5}\whtd1100{#6}%
       \whtd1100{#7}\whtd1000{#8}} }$}}

\def\eddeiii#1.#2.#3.#4.#5.#6.#7.#8.#9.{\hbox{$\vcenter{\hbox
       {\dynk \whtd0100{#1}\whtd1100{#3}%
       \up1{\whtr0001{#2}}\whtd1110{#4}\whtd1100{#5}\whtd1100{#6}%
       \whtd1100{#7}\whtd1100{#8}\whtd1000{#9}} }$}}

\baselineskip=17pt 

\section{Introduction}
\setcounter{equation}{0}

The affine Toda field theory is one of the best understood field theories
at the classical \cite{MOPa}\ts and at the quantum levels [2--8],
thanks to its integrability.
It is the close connection between the affine Toda field theory and
the conformal field theory in 2 dimensions,
another group of best understood quantum field theories,
 (integrable deformation of conformal field theory \cite{ZEYSY,HMBL})
that led to the interesting but controversial ``imaginary coupling" affine
Toda field theory.

Toda field theory is integrable at the classical level due to
the presence of an infinite number of conserved quantities.
Many beautiful properties of Toda field theory, both at the classical
and quantum levels, have been uncovered in recent years.
In particular, it is firmly believed that the integrability survives
quantisation.
The exact quantum S-matrices are known for all the
Toda field theories based on non-simply laced algebras \cite{DGZc,CDS}
as well as those
based on simply laced algebras [2--6].
The singularity structure of the latter S-matrices, which in some
cases contain poles up to 12-th order \cite{BCDSc},
is  beautifully explained in terms of the
singularities of the corresponding Feynman diagrams
\cite{BCDSe}, so called Landau singularities.

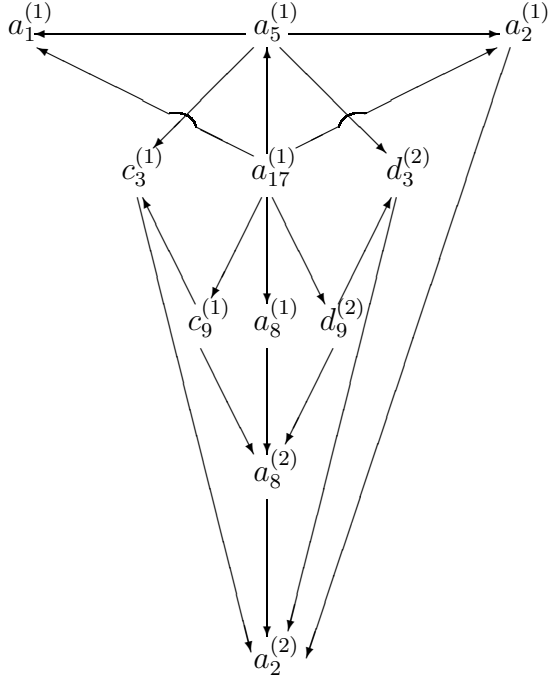
\begin{figure}
\begin{picture}(400,280)
\put(199,210){$a_{17}^{(1)}$}
\put(205,203){\vector(0,-1){40}}
\put(205,220){\vector(0,1){40}}
\put(207,203){\vector(1,-2){20}}
\put(203,203){\vector(-1,-2){19}}
\put(225,153){$d_9^{(2)}$}
\put(200,153){$a_8^{(1)}$}
\put(175,153){$c_9^{(1)}$}
\put(205,146){\vector(0,-1){40}}
\put(230,146){\vector(-1,-2){18}}
\put(180,146){\vector(1,-2){20}}
\put(200,96){$a_8^{(2)}$}
\put(205,89){\vector(0,-1){52}}
\put(200,25){$a_2^{(2)}$}
\put(232,163){\vector(1,2){20}}
\put(178,163){\vector(-1,2){20}}
\put(200,265){$a_{5}^{(1)}$}
\put(210,260){\vector(1,-1){40}}
\put(200,260){\vector(-1,-1){38}}
\put(150,210){$c_{3}^{(1)}$}
\put(250,210){$d_{3}^{(2)}$}
\put(197,265){\vector(-1,0){80}}
\put(213,265){\vector(1,0){80}}
\put(295,265){$a_{2}^{(1)}$}
\put(107,265){$a_{1}^{(1)}$}
\put(156,203){\vector(1,-4){43}}
\put(254,203){\vector(-1,-4){41}}
\put(297,260){\vector(-1,-3){77}}
\put(216,222){\line(2,1){16}}
\put(200,219){\line(-2,1){22}}
\put(242,235){\vector(2,1){50}}
\put(168,235){\vector(-2,1){50}}
\qbezier(232,230)(234,236)(242,235)
\qbezier(178,230)(176,236)(168,235)
\end{picture}
\caption{Reductions of $a_{17}^{(1)}$ Toda Theory}
\label{ared}
\end{figure}

In this present note,
 we are mainly concerned with the various types of classical
solutions of the affine Toda field theory.
The structure of the classical equation of motion is common
to both the real and imaginary coupling regimes. They share an infinite
set of conserved quantities.
To be more specific, we discuss various types of classical equations
derived by the \lq reduction' (see Olive and Turok
\cite{OT}) from a given affine Toda field theory.
As is well known, the reduction of affine Toda field theory by
keeping its integrability is closely related to the Dynkin diagram
automorphisms of the underlying Lie algebras.
Reductions of special interest are the ``dimension one'' reductions
\cite{Sas},
in which a single field degree of freedom satisfies the entire set of
equations for the $r$-component field ($r$ is the rank of the
underlying algebra). The situation is close to the classical images of
the soliton, which can exist by itself, thus satisfying a
single-component nonlinear equation of motion.
However, the affine Toda field theory in the real coupling regime
has no \Sols because of the unique vacuum.
In the imaginary coupling regime the potential becomes periodic
and the possibility of \Sols emerges ( although non-hermitian in
most cases).
In such a situation the solutions of the \Do reduced equations are
expected to
play important r\^oles just as the \Sols and \Brs in the sine-Gordon
theory. It is also hoped that an understanding of
the imaginary coupling regime would
give us some hints for the real coupling regime,
as well as for the foundation of the quantum group structure.
Here we study in detail the possible
reductions of affine Toda theories. About a decade ago Olive and
Turok \cite{OT}, addressed the same problem. Subsequently a few years
ago one of the authors studied the problem \cite{Sas}, concentrating
mainly on ``dimension one'' reduction, and obtained some new
results. In the following pages we follow the footsteps of these
papers and investigate many new reductions which were not mentioned
in the earlier papers. Along with earlier results we will try to give a
complete and exhaustive list of all possible affine Toda reductions.
We shall
also present the general rules for obtaining the reductions.
Furthermore, some diagrammatic techniques will be presented to obtain
reductions of some particular theories. These diagrams will help to
figure out which of the roots of the Dynkin diagrams are to be
identified. In the ref.\cite{OT} two types of reductions were
mentioned viz. non-direct and direct depending on whether or not the
two adjacent points of Dynkin diagram are identified. It was mentioned
that in case of non-direct reduction one has to rescale the
space-time(or the mass parameter) to obtain the
proper reduction. But we find that this is not always the case. In the
ref.\cite{OT} only the symmetries of
simply laced algebras were considered and subsequently  folded to
obtain the non-simply laced algebras and reductions. The folding of
non-simply laced algebras were ignored because the didn't produce new
algebras. In the present paper we also consider the symmetries and the
folding of the non-simply laced algebras. Although they
don't provide any new algebra but the Toda equations based on them
reduce to other Toda theories based on  non-simply laced (and
sometimes simply laced) algebras of
lower rank. We also
hope that these reductions will provide a better ground for the study of
the affine Lie algebra. We think that just by looking at the Dynkin
diagram of some affine Lie algebra one may not have a hint for
the real number of reductions of Toda theory based on that. We hope
the following diagram will illustrate the above point. In Fig. \ref{ared}
we list the all possible reductions of $a_{17}^{(1)}$ theory.

The organisation of the present note is as follows.
A very brief summary of the rudimentary facts of the affine Toda field
theory is given in the next section. In  section 3 we give the
general reduction procedure. The subsequent sections  discuss case by
case in detail with diagrammatic techniques.

\section{Summary: Affine Toda Field Theory }

In this section we would like to
recapitulate the essential features of Toda field theories which we
believe is  helpful to understand the problem.
Affine Toda field
theory \cite{MOPa} is a
massive scalar field theory with exponential interactions in $1+1$
dimensions
described by the Lagrangian
\begin{equation}
{\cal L}={1\over 2}
\partial_\mu\phi^a\partial^\mu\phi^a-V(\phi ),
\label{ltoda}
\end{equation}
in which the potential is given by
\begin{equation}
V(\phi )={m^2\over
\beta^2}\sum_0^rn_je^{\beta\alpha_j\cdot\phi}.
\label{vtoda}
\end{equation}
The field $\phi$ is an $r$-component scalar field, $r$ is the rank of a
compact semi-simple Lie algebra $g$ with $\alpha_j$;
$j=1,\ldots,r$ being its simple roots. The roots are normalised so
that long roots have length 2, $\alpha_L^2=2$. An additional root,
$\alpha_0=-\sum_1^rn_j\alpha_j$ is an integer linear combination of
the simple roots, is called the affine root;
 it corresponds to the extra spot on an extended Dynkin diagram
for $\hat g$ and $n_0=1$. We call the integers $n_i$ the `weight',
which are sometimes referred to as Kac's labels, too.
When the term containing the extra root is removed, the theory becomes
conformally invariant (conformal Toda field theory). Then the theory
is based on the root system of a finite Lie algebra `$g$' and sometimes
it is called a non-affine Toda theory in distinction with the affine one.
The simplest affine Toda field theory, based on the simplest Lie algebra
$a_1^{(1)}$, the algebra of ${\wh su(2)}$,
is called sinh-Gordon theory, a cousin of
the well known sine-Gordon theory. $m$ is a real parameter setting
the mass scale of the theory and $\beta$ is a real coupling constant,
which is relevant only in quantum theory.
Since the coupling constant is  irrelevant
for the discussion of the classical solutions, we
put $\beta=1$ hereafter.
The equation of motion reads (see \cite{BSa} for various forms of
affine Toda equation of motion)
\beq
 \partial\sp2\phi = -m^2\sum_{j=0}\sp r n_j\alpha_je\sp{\alpha_j
\cdot\phi} . \label{eom}
\eeq

It turns out that the data in quantum theory,
such as the masses and couplings
of various kinds, are also useful for the reduction of
classical equation of motion. Expanding the potential part
\rref{vtoda}\
of the Lagrangian up to second order, we find
\beq
V({\bf \phi} )=m^2\left(\sum\sp r_{i=0}n_i +
{1\over2}\sum\sp r_{i=0}n_i\alpha_i\sp a\alpha_i\sp b\phi\sp a\phi\sp b
+\dots \right) ,
\label{lagexp}\eeq
from which we can extract a $({\rm mass})\sp2$ matrix
\beq
h=\sum\sp r_0n_i, \quad
(M\sp2)\sp{ab}=m^2\sum\sp r_0n_i\alpha_i\sp a\alpha_i\sp b .
\label{mass}\eeq
Here $h$ is the Coxeter number \cite{Mb,KMb}.
The mass matrix has been studied before \cite{BCDSc,MOPa,BCDSb,CMa,KMb}.

One important fact which underlies the present work is that
the particles of the simply laced theory
are associated unambiguously with the spots on the Dynkin diagram
and thus to the simple roots (fundamental weights) of
the associated finite Lie algebra
\cite{BCDSc,FrFLO,BCDSb,FKM}.
It is based on the
observation that the set of masses computed
as the $r$ eigenvalues of the mass matrix \rref{mass} actually
constitute the Frobenius-Perron eigenvector of the Cartan matrix of
the associated finite Lie algebra. In other words, if we set
${\bf m}=(m_1,m_2,\dots ,m_r)$
then
\beq
C{\bf m}=\lambda_{\rm min}{\bf m}=4m^2\sin\sp2{\pi\over 2h}\ {\bf m}
\label{fp}
\eeq
where $C$ is the Cartan matrix
$C_{ij}=2\alpha_i\cdot\alpha_j/\alpha_j\sp2$,
$i,j=1,\dots ,r. $
The Coxeter numbers and $({\rm mass})\sp2$ of various theories
together with the Dynkin diagrams and particle labelling,
can be found in ref. \cite{BCDSc}.

The idea of folding and reductions based on the symmetry
(automorphism) of the Dynkin diagram is simple: a symmetry of the Dynkin
diagram, permuting the
points as $\alpha\rightarrow p(\alpha)$, can be rewritten as a mapping
of the
field space to itself, $\phi\rightarrow p(\phi)$. This is a symmetry of
the classical field equations derived from the Lagrangian with the
potential \rref{vtoda}, namely it maps a solution to another.
This means that if the fields initially take
values in the subspace invariant under $p$, they will remain there,
at least classically. Since the subspace is of smaller dimension than
the original field space, the evolution of fields within it can be
described in terms of an equation with fewer variables than the
original equation. The latter
is obtained by projecting the variables $\alpha_i$ in eqn. \rref{vtoda}
onto the invariant subspace. This process of obtaining new equations
and their solutions from the old, by exploiting diagram symmetries,
is known as reduction.
In other words, an arbitrary solution of a reduced theory always gives
solution(s) of the original theory by appropriate embedding(s).
The so-called direct reductions are those such that $\alpha\cdot p(\alpha
)=0$ for each root $\alpha$ (i.e. the symmetry does not relate
points linked by a line on the Dynkin diagram).

One further point. A symmetry of the unextended Dynkin diagram of a
simply-laced algebra yields the diagram for one of the non simply-laced
algebras on projection onto the invariant subspace, and the addition of
the extra point to extend the diagram always respects such a symmetry.
The resulting projected diagram is the untwisted affine diagram for
the non
simply-laced algebra. This is a reflection of the fact that the symmetry
group of the extended diagram contains always at least that of the
unextended diagram.
Reductions involving any
additional symmetries of the extended diagram yield affine Toda theories
based on the twisted affine Dynkin diagrams.
The reductions based on the symmetries of the unextended Dynkin
diagrams are discussed rather completely in earlier papers
\cite{BCDSc,OT,Sas}, therefore in this note we concentrate on the
reductions based on the symmetries of the affine Dynkin diagrams.
There is an interesting distinction to be made
here between the two different types of non simply-laced theories,
namely twisted and untwisted theories.
The foldings leading to untwisted theories turn out to remove
degeneracies from the mass spectrum, the resulting non degenerate
particles always being linear combinations of the degenerate particles
in the parent theory (it is precisely this \lq rediagonalisation'
which causes the problems in the quantum theory
mentioned above). In contrast, foldings leading to twisted diagrams
remove some particles from the spectrum altogether, while leaving
the others unchanged.

Quantum S-matrices of all affine Toda theories are also known
\cite{AFZa,BCDSa,BCDSc,CMa,DDa,CKM}.
Based on the assumption that the infinite set of conserved quantities
be preserved after quantisation, only the elastic processes are
allowed and the multiparticle S-matrices are factorised into
a product of two particle elastic S-matrices.

\section {Reduction}
\setcounter{equation}{0}
In this section we
show mathematically  what we mean actually by Toda reduction,
stated in earlier sections.
Consider the Toda equation of motion (\ref{eom}) for the algebra `$g$'
of rank `$r$'
\beq
 \partial\sp2\phi = -m^2\sum_{\alpha_j\in g,~j=0}\sp r n_j\alpha_j
e\sp{\alpha_j \cdot\phi} . \label{eoma}
\eeq
Now let us make the following special ansatz for the field $\phi$,
\beq
        \phi=\sum_{i=1}^p\beta_i\varphi_i, \qquad p<r, \label{veom}
\eeq
in which $\beta_i$'s are $p$ constant vectors constructed
with linear combinations of $\alpha_j$'s that is $\beta_i=\sum_j
a_{ij}\alpha_j$ ($a_{ij}$'s constant), and
$\varphi$ is a $p$-component real field.
In other words \rref{veom} simply means that the solution $\phi$ lies
in the $p$-dimensional subspace spanned by $\bet{i}$'s. The explicit
parametrisation $\varphi_i$ is immaterial as in the original \rref{eoma}.
Now inserting (\ref{veom}) in the eqn. (\ref{eoma}), we have
\beq
\sum_{1}^p\bet{i}\partial\sp2\varphi_i = -m^2\sum_{\alpha_j\in
g,~j=0}\sp r n_j\alpha_j
e\sp{\alpha_j \cdot\sum_i\beta_i\varphi_i} . \label{eomb}
\eeq
Now if $\beta_i$'s behave like the roots of rank `$p$' algebra
`$g'$' and if the equation (\ref{eomb}) can be recast as
\beq
 \partial\sp2\phi = -{m'}^2\sum_{\beta_i\in g',~i=0}\sp
p {n'}_i\beta_i
e\sp{\beta_i \cdot\phi}, \label{eomc}
\eeq
we will succeed to obtain the reduction\footnote[1]{There are cases
where the $\bet{i}$'s behave like the roots of a lower rank algebra but
Toda equation would not retain its form as eqn. \rref{eoma}. For
example if we start with \cno theory and identify only two end
vertices i.e. $\alp0$ and $\alp{n}$ to construct $\bet{n-1}$, and leave
every other root as it is, so that $\bet{i}=\alp{i+1},~ i=0,1,..,n-2$,
and $\bet{n-1}={1\over 2}(\alp0+\alp{n})$, we find that $\bet{i}$'s do
indeed behave like roots of $a_{n-1}^{(1)}$ theory, but in this case
Toda equation based on algebra \cno does not reduce to one based on
$a_{n-1}^{(1)}$.}. In the above equation
${n'}_i$ are `weights' for the algebra $g'$, and we have
$\sum_{i=0}^p {n'}_i\beta_i=0$. Note due to the choice of the ansatz
(\ref{veom}) the field $\phi$ appearing
in the above equation is a $p$ component
field. In most of the cases the mass parameter $m^2={m'}^2$. But in
some cases a constant scaling factor is needed to relate these two mass
parameters (We shall comment on this point at the end of the next section).

Now we illustrate the above procedure with an example. Consider the
reduction of Toda equation based on affine algebra $f^{(1)}_{4}$ to
one based on algebra $d_4^{(3)}$,
The Dynkin diagram for $f_4^{(1)}$ (now on we follow the notations of
the ref. \cite{H}) is
\beq
\eddfid{\alpha_0}.{\alpha_1}.{\alpha_2}.{\alpha_3}.
{\alpha_4}.\llap{\eddfiu1.2.3.4.2.},
\label{ffon}
\eeq
in which we show $n_i$ on each simple root.
Now we construct two vectors $\beta_1$ and $\beta_2$ in the following
way
\beq
    \beta_1={1\over 3}(\alpha_1+2\alpha_3),\qquad \bet2=\alp2
\label{droo}\eeq
so that $\beta_1$ and $\beta_2$ behave like the roots of affine
algebra $d_4^{(3)}$(see also the last diagram of Fig.\rref{fig:dab}).
 The affine root $\beta_0$ can be obtained using
relations $\sum_{i=0}^2 {n'}_i\beta_i=\beta_0+2\beta_1+\beta_2=0$
and $\sum_{j=0}^4n_j\alp{j}=0$, giving,
\beq
\beta_0={1\over3}(\alpha_0+2\alpha_4),
\label{drooa}
\eeq
which is equivalent to identifying $\alp0$ with $\alp4$ in the diagram
\rref{ffon}.
The various inner products of the roots of \dfth are following:
\beq
\bet0^2=\bet1^2={2\over3},~\bet2^2=2,~\bet0\cdot\bet1=-{1\over3},
{}~\bet1\cdot\bet2=-1,~{\rm and}~\bet0\cdot\bet2=0.
\label{dotpb}
\eeq
In this case no re-normalisation of the roots is necessary because the
longest root is already having squared length of 2 units. We now
tabulate various inner products of $\alp{i}$'s and $\bet{i}$'s
\Bear
\alp0\cdot\bet1=-{1\over 3}, & \alp0\cdot\bet2=0,\nonumber \\
\alp1\cdot\bet1={2\over 3}, & \alp1\cdot\bet2=-1,\nonumber \\
\alp2\cdot\bet1=-1, & \alp2\cdot\bet2=2,\nonumber \\
\alp3\cdot\bet1={2\over 3}, & \alp3\cdot\bet2=-1,\nonumber \\
\alp4\cdot\bet1=-{1\over 3}, & \alp4\cdot\bet2=0.
\label{dotpr}
\Enar
Now we start with the equation \rref{eoma} for the algebra
$g\equiv$\ffo. Equation \rref{eoma} looks in the expanded form:
\beq
\partial^2\phi=-m^2(\alp0e^{\alp0\cdot\phi}+2\alp1e^{\alp1\cdot\phi}+
3\alp2e^{\alp2\cdot\phi}+4\alp3e^{\alp3\cdot\phi}+
2\alp4e^{\alp4\cdot\phi}).
\label{expa}
\eeq
Furthermore we take $\phi=\beta_1\varphi_1+\beta_2\varphi_2$, and use
\rref{dotpr} in eqn. \rref{expa} to obtain,
\beq
\bet1\partial^2\varphi_1+\bet2\partial^2\varphi_2=-m^2[
(\alp0+2\alp4)e^{-{1\over3}\varphi_1}+
2(\alp1+2\alp3)e^{{2\over3}\varphi_1-\varphi_2}
+3\alp2e^{-\varphi_1+2\varphi_2}]\label{expb}
\eeq
using \rref{dotpb},\rref{droo} and \rref{drooa}
in above equation, we find that it can be
rewritten as,
\beq
\partial^2\phi=-3m^2(\bet0e^{\bet0\cdot\phi}+2\bet1e^{\bet1\cdot\phi}+
\bet2e^{\bet2\cdot\phi})
\eeq
or in compact form,
\beq
 \partial\sp2\phi = -{m'}^2\sum_{\beta_j\in d_{4,}^{(3)}~j=0}\sp2
{n'}_j\beta_je\sp{\beta_j \cdot\phi} . \label{eomd}
\eeq
In this case we find that ${m'}^2=3m^2$ and ${n'}_j$'s are appropriate
`weights' of the $d_4^{(3)}$ algebra.

At this stage we would like to make some interesting observations. At
first sight it appears that the diagram \rref{ffon} does not have any
symmetry at all and therefore it cannot be folded. But if one pays a
little attention one observes that indeed a kind of symmetry is
present in the diagram \rref{ffon}. The vertex pair $(\alp0,\alp1)$ is
more or less same like the pair $(\alp4,\alp3)$. The two vertices in
each pair are of the same length, and their positions from the
center (i.e. from vertex $\alp2$) are the same. The pair
$(\alp0,\alp1)$ is
joined to the vertex $\alp2$ with a single line whereas the pair
$(\alp4,\alp3)$ is joined to $\alp2$ with a pair of lines, but the
`weights' (Kac's labels) of the latter pair is just twice of that of the
former one. This is why we identify the vertices $\alp1$ with
$\alp3$ and $\alp0$ with $\alp4$ and give the twice weightage to $\alp3$
and $\alp4$ (see equations \rref{droo},\rref{drooa}). To see the
symmetry in
a more picturesque way we would like to draw a ``reduced diagram'' from
\rref{ffon} by dividing the `weights' of the vertices (pointed by the
arrow in the diagram) by the number of lines joining them from the
center, and replacing the multiple lace with the single lace in the
diagram. In this case there are two lines joining the pair
$(\alp3,\alp4)$(arrow points these two vertices) with the vertex $\alp2$.
Their original `weights' 4 and 2 are divided by two giving new `weights'
equal to 2 and 1 respectively. Furthermore we remove the double lace from
the diagram \rref{ffon} to have the following symmetric ``reduced diagram''
with the `weights' shown
\beq
\andfive1.2.3.2.1. .
\eeq
For further illustration consider another diagram \gto, which does not
seem to have any symmetry at first sight,
\beq
\eddgid{\alp0}.{\alp1}.{\alp2}.\llap{\eddgiu{1}.{2}.{3}.}\eeq
In this case $\alp2$ is joined to $\alp1$ with three lines and the
arrow is
pointing $\alp2$, so we divide the `weight' of $\alp2$ by three and
replace three lines by a single line resulting the following symmetric
``reduced diagram'' with the new `weights', as
\beq
\andthr1.2.1.
\eeq
And we find that \gto can be folded into $a_2^{(2)}$, by identifying
$\alp0$ with $\alp2$ and three times weightage to $\alp2$.

Now we are in a position to give the rules for the folding.
The first thing is to observe whether the Coxeter number,
$h=\sum_{i=0}^rn_i$ of the algebra in
question (i.e. on which the Toda theory one wants to reduce is based)
is a prime or not.

\begin{description}
\item[Case a)] $h$ is a prime. Now we check whether the algebra
         is $a_{2n}^{(2)}$. Here again there are two possibilities.
\begin{enumerate}
 \item The algebra is $a_{2n}^{(2)}$. In this case there is no
      reduction possible.
 \item  The algebra is one of \ano, \dnt, and \aont. In all the cases the
      theory will reduce to $a_{2m}^{(2)}$ which is having lower
     rank than that  of the original one (i.e. $m<n$).
\end{enumerate}
\item[Case b)]$h$ is not a prime. There are three  cases:
  \begin{enumerate}
  \item Theory is based on an exceptional algebra ie on \esio, \eseo
       or \eeho. Very limited number of reductions are possible and
       are given in section 8.
    \item Theory is based on \dfth algebra. No reduction is possible.
     \item Theory is not any of above two cases.
     In this case we find all the divisors of
    the $h$. Corresponding to  each divisor we have one or more
    reductions. By this we mean the reduced theory will be based on an
    algebra which will have the Coxeter number equal to the divisor.
    Regarding the type of reductions possible we have to  refer
    the following sections where everything will be treated case by case.
    Next go back to the beginning of the programme to check whether
    these reduced theories can be reduced further or not. All the
    theories will finally reduce to the \atnt\ theories with $2n+1$
    prime or $a_1^{(1)}$ with the exception
    of \ffo\ which reduces to \dfth.

\end{enumerate}
\end{description}

Now we will present rules for the folding of the Dynkin diagram.
The detailed diagrams are given in the subsequent sections. When we
identify the $p$-points in the Dynkin diagram say corresponding
to the roots $\alp{i_1}$, $\alp{i_2}$, $\alp{i_3}$, $\cdots$,
$\alp{i_p}$(with `weights' $n_{i_1}$, $n_{i_2}$, $n_{i_3}$, $\cdots$,
$n_{i_p}$, respectively), we construct
a root, $\bet{i}$ for the new algebra as the following weighted average of
$\alp{i_j}$'s,
\beq
  \bet{i}={1\over{\sum_{j=1}^p n_{i_j}}}{\sum_{j=1}^p n_{i_j}\alp{i_j}}.
\label{wetav}
\eeq
The new roots, $\bet{i}$'s, behave like the roots of the
reduced algebra, but it is possible that the long root(s) may not have
the squared length of 2. In that case one has to re-normalise all the root
such that the long roots have squared length of 2 units.

\section{Folding of Dynkin diagrams- \ano\ series}
\setcounter{equation}{0}
In this and the following sections we will draw the Dynkin diagrams in a
convenient way so that folding is achieved easily.
In all the following diagrams the vertices falling on the same column are
to be identified and new roots are constructed using the expression
\rref{wetav}. In all the diagrams vertices are labelled by the
roots(for the corresponding `weights', $n_i$'s, see Ref. \cite{H}).

There are roughly two types of the reduction. In the first case the theory
reduces to one of the members of the same series, and in the second
case it reduces to some member of the other series. We first discuss the
former case.
The reduction of the \ano\ affine Toda field theory is quite
straightforward
\cite{BCDSc,OT,Sas}
and has a rich structure, which reflects the
$Z_{n+1}$ symmetry of the extended Dynkin diagram.

If the Coxeter number $h=n+1$ is not a prime i.e. $n+1 =pq $, for
integers $p$, $q$ then the above equation (\ref{eom}) can be
reduced to $a_{p-1}\sp{(1)}$  and $a_{q-1}\sp{(1)}$ Toda theories
by dividing by $Z_q$ and
$Z_p$ , respectively:
\Bear a_{pq-1}\sp{(1)} \rightarrow
a_{p-1}\sp{(1)}, \qquad
{\rm divided \ by}\quad
Z_q ,     \\
a_{q-1}\sp{(1)}, \qquad
{\rm divided \ by}\quad Z_p .
\label{pqred}
\Enar
The reductions listed above are so-called direct reductions, i.e.,
the roots to be identified (folded) are not connected by a link in
the Dynkin diagram.
For details we refer to  a previous paper
\cite{BCDSc}.

We construct the new roots for the reduced theory in the following way
for the above reductions:
\beq
\bet{i}={1\over\sqrt{p}}\left[\sum_{j=0}^{p-1}\alp{i+jq}\right]~~;i=1,2,
\cdots,q-1;\qquad({ a_{pq-1}^{(1)}\rightarrow a_{q-1}^{(1)}}),\eeq
\beq
\bet{i}={1\over\sqrt{q}}\left[\sum_{j=0}^{q-1}\alp{i+jp}\right]~~;i=1,2,
\cdots,p-1;\qquad({a_{pq-1}^{(1)}\rightarrow a_{p-1}^{(1)}}).\eeq

In the second case
$a_n\sp{(1)}$ theory may reduce to one of the following theories namely
those based on non simply-laced theories, see
Figs. \rref{aotc},\rref{aotd} and \rref{aoat}.

\Bear
a_{2n-1}\sp{(1)} \rightarrow c_n\sp{(1)}\\
a_{2n+1}\sp{(1)} \rightarrow d_{n+1}\sp{(2)}\\
a_{2n}\sp{(1)} \rightarrow a_{2n}\sp{(2)}
\label{Cnred}
\Enar

First of these, (see Fig. \ref{aotc})
is an example of what is often called ``direct-reduction'' in early
literature. In this case
the vertices identified have the same masses. So the reduced theory has
the same masses of the parent theory having the degeneracy removed.
In this case no two neighbouring vertices are identified.
In the second one, expression (4.6), Fig. \ref{aotd}, the vertices
identified have different masses and at the ends the the neighbouring
vertices are identified. Finally in the last one (see
Figure \ref{aoat}), one identifies the vertices with the same masses.
An interesting point to note is that any other rotation of the parent
theory will give these reductions. So there are $n$ ways to obtain
first two of these reductions and $2n+1$ ways to get the last one.
But in all these rotated case the masses of the identified vertices
will be different. Here we make some comments on the `direct' and
`non-direct' reductions. In ref. \cite{OT} the distinction between
these two reductions was made by following two points. First, in case of
`non-direct' reductions one identifies vertices which are directly
linked in the Dynkin diagram, and second, one needs a rescaling of
space-time (or mass parameter) to obtain the reduction.
But we find that these two points don't go hand in hand. In the case of
the reduction \ffo$\rightarrow$\dfth (see earlier sections), two
directly linked vertices were never identified, but one needed a mass
parameter rescaling, whereas in the last two reductions, Figs. \rref{aotd}
and \rref{aoat}, the directly linked vertices were identified but the
mass rescaling was not needed. So, we find that the distinction becomes
obscure.

\section {\dnt~ and~ \cno~ theories}
\setcounter{equation}{0}
The reductions of \dnt\ and \cno\ theories are very similar due to the
nature of their affine Dynkin diagrams. The Dynkin diagram of the one can
be obtained just by reversing the arrows in the diagram of the
other. In other words these two Dynkin diagrams are {\it dual\/} to each
other.
The reductions can be obtained with the help of the
Figs. \ref{fig:dtd} and \ref{fig:dcao}. These figures guide to construct
the new roots of the reduced theory. In all the cases vertices falling
on the same vertical line are identified.
There are three types of reductions.

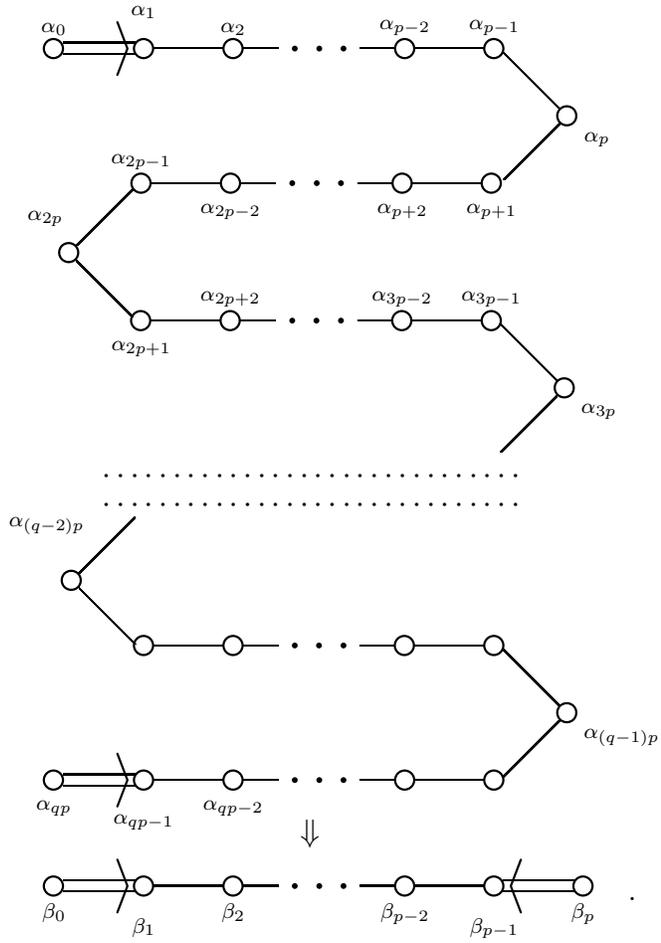
\begin{figure}
\begin{picture}(300,370)
\put(225,360){$q$~even}
\put(142,322){$\ncdfrdc{\alp{0}}.{\alp{1}}.{\alp{2}}.
                 {\alp{p-2}}.{\alp{p-1}}.{\alp{p}}.$}
\put(144,258){$\ncdfal{\alp{2p}}.{\alp{2p-1}}.{\alp{2p-2}}.
     {\alp{p+2}}.{\alp{p+1}}.$}
\put(175,219){$\ncdfar{\alp{2p+1}}.{\alp{2p+2}}.
     {\alp{3p-2}}.{\alp{3p-1}}.{\alp{3p}}.$}
\put(177,162.5){$\cdots\cdots\cdots\cdots\cdots\cdots%
         \cdots\cdots\cdots\cdots$}
\put(177,151.5){$\cdots\cdots\cdots\cdots\cdots\cdots\cdots%
             \cdots\cdots\cdots$}
\put(145,134){$\lronit{\alp{(q-2)p}}.$}
\put(176,96){$\ncdfar{}.{}.{}.{}.{\alp{(q-1)p}}.$}
\put(142,45){$\ncdddc{\alp{qp}}.{\alp{qp-1}}.{\alp{qp-2}}.{}.{}.$}
\put(252,27){$\Downarrow$}
\put(142,5){$\ddcnds{\bet{0}}.{\bet{1}}.{\bet{2}}.
    {\bet{p-2}}. {\bet{p-1}}.{\bet{p}}.$}
\end{picture}
\caption{$c_{pq}^{(1)}\rightarrow c_{p}^{(1)}(c_q^{(1)})$}
\label{fig:ctce}
\end{figure}

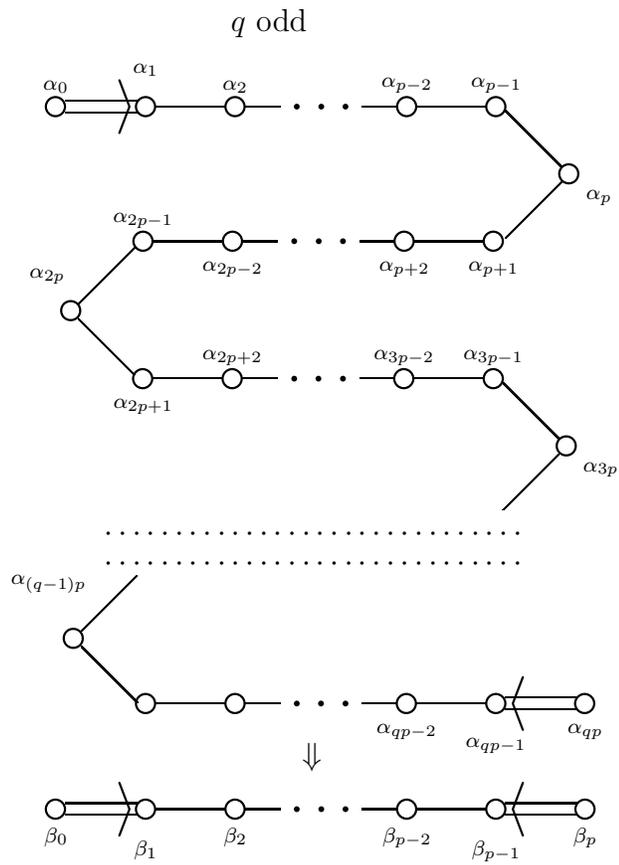
\begin{figure}
\begin{picture}(300,315)
\put(225,305){$q$~odd}
\put(142,271){$\ncdfrdc{\alp{0}}.{\alp{1}}.{\alp{2}}.
            {\alp{p-2}}.{\alp{p-1}}.{\alp{p}}.$}
\put(144,207){$\ncdfal{\alp{2p}}.{\alp{2p-1}}.{\alp{2p-2}}.
     {\alp{p+2}}.{\alp{p+1}}.$}
\put(175,168){$\ncdfar{\alp{2p+1}}.{\alp{2p+2}}.
     {\alp{3p-2}}.{\alp{3p-1}}.{\alp{3p}}.$}
\put(177,111.5){$\cdots\cdots\cdots\cdots\cdots\cdots\cdots%
       \cdots\cdots\cdots$}
\put(177,100.5){$\cdots\cdots\cdots\cdots\cdots\cdots\cdots%
        \cdots\cdots\cdots$}
\put(145,83){$\lronit{\alp{(q-1)p}}.$}
\put(176,45){$\ncddco{}.{}.{\alp{qp-2}}.
     {\alp{qp-1}}.{\alp{qp}}.$}
\put(252,27){$\Downarrow$}
\put(142,5){$\ddcnds{\bet{0}}.{\bet{1}}.{\bet{2}}.
    {\bet{p-2}}. {\bet{p-1}}.{\bet{p}}.$}
\end{picture}
\caption{$c_{pq}^{(1)}\rightarrow c_{p}^{(1)}(c_q^{(1)})$}
\label{fig:ctco}
\end{figure}

The first type is where $d_{pq}^{(2)}(c_{pq}^{(1)})$ theories reduce
to $d_p^{(2)}(c_p^{(1)})$ or $d_q^{(2)}(c_q^{(1)})$ (see
Figs. \ref{fig:dtd},\ref{fig:ctce} and \ref{fig:ctco}).
Although the reduction type is similar for \dnt\
and \cno\ theories, the diagrams for these reductions have some
differences. Notice that in the case of \dnt\ theories on the edges one
identifies two adjacent vertices of Dynkin diagram Fig. \ref{fig:dtd},
whereas for \cno\ theories we have
single vertex on the edges, Figs. \ref{fig:ctce} and \ref{fig:ctco}.

In the second type of reduction $d_{2q}^{(2)}(c_q^{(1)})$ theories go to
\ano\ theory. The construction of a new root $\bet1$ for $a_1^{(1)}$ is
depicted in Fig.\ts\ref{fig:dcao}. In this case diagrams are similar
to that of the above
type but there are only two columns.

Lastly $d_{p(2n+1)}^{(2)}(c_{p(2n+1)}^{(1)})$ reduce to \atnt\
theories Figs.\ts\ref{fig:dat} and \ref{fig:cat}. In this case on one
edge of the diagram
two neighbouring vertices are identified whereas the other edge has
a single vertex.

\section{\atnt\ series}
\begin{figure}
\begin{picture}(300,325)
\put(142,313){$\ncdfrd{\alp{0}}.{\alp{1}}.{\alp{2}}.
     {\alp{n-2}}.{\alp{n-1}}.{\alp{n}}.$}
\put(142,262){$\ncddld{\alp{2n}}.{\alp{2n-1}}.{\alp{2n-2}}.
     {\alp{n+2}}.{\alp{n+1}}.$}
\put(142,211){$\ncdfrdl{\alp{2n+1}}.{\alp{2n+2}}.{\alp{2n+3}}.
     {\alp{3n-1}}.{\alp{3n}}.{\alp{3n+1}}.$}
\put(162,153.5){$\cdots\cdots\cdots\cdots\cdots\cdots\cdots%
      \cdots\cdots\cdots$}
\put(162,140){$\cdots\cdots\cdots\cdots\cdots\cdots\cdots%
      \cdots\cdots\cdots$}
\put(142,96){$\ncdfrdl{\alp{(p-1)(2n+1)}}.{}.{}.{}.
      {}.{\alp{(p-1)(2n+1)+n}}.$}
\put(142,45){$\ncdded{\alp{p(2n+1)-1}}.{}.{}.{}.{}.$}
\put(252,27){$\Downarrow$}
\put(142,5){$\datwono{\bet{0}}.{\bet{1}}.{\bet{2}}.
    {\bet{n-2}}. {\bet{n-1}}.{\bet{n}}.$}
\end{picture}
\caption{$d_{p(2n+1)}^{(2)}\rightarrow$ \atnt} \label{fig:dat}
\end{figure}
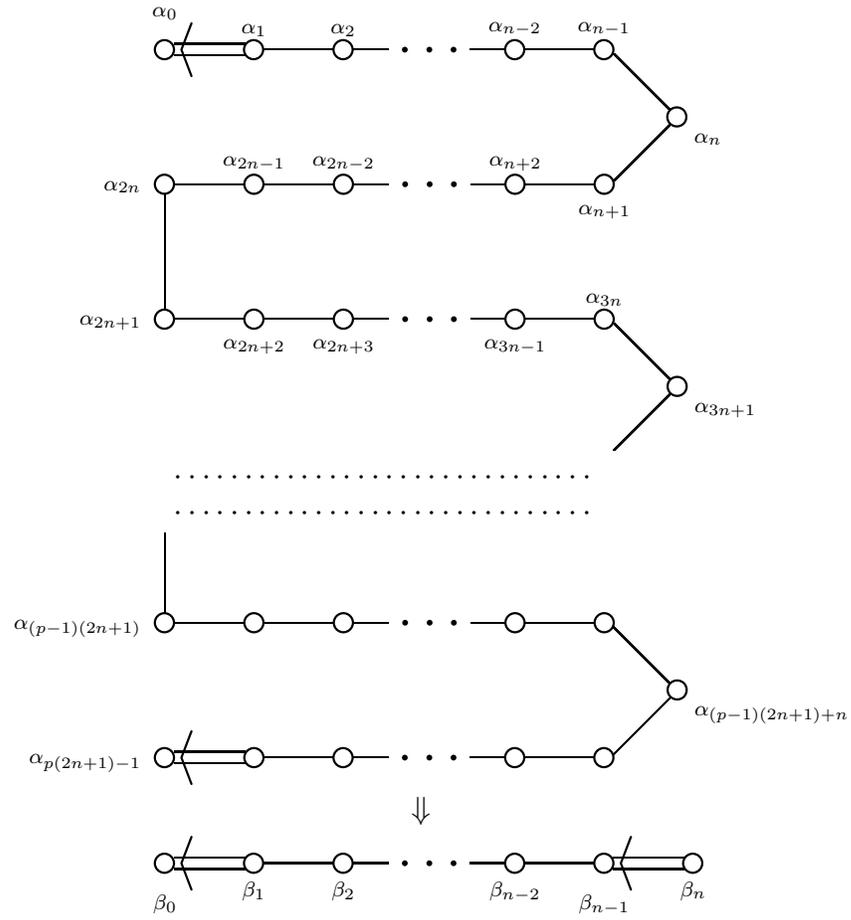

\begin{figure}
\begin{picture}(300,325)
\put(142,322){$\ncddcd{\alp{0}}.{\alp{1}}.{\alp{2}}.
                 {\alp{n-1}}.{\alp{n}}.$}
\put(145,258){$\ncdfalu{\alp{2n+1}}.{\alp{2n}}.{\alp{2n-1}}.
     {\alp{n+2}}.{\alp{n+1}}.$}
\put(176,219){$\ddanufd{\alp{2n+2}}.
     {\alp{2n+3}}.{\alp{3n}}.{\alp{3n+1}}.$}
\put(165,169){$\cdots\cdots\cdots\cdots\cdots\cdots\cdots%
              \cdots\cdots\cdots$}
\put(165,158){$\cdots\cdots\cdots\cdots\cdots\cdots\cdots%
            \cdots\cdots\cdots$}
\put(145,96){$\ncdfalu{\alp{p(2n+1)}}.{}.{}.
     {}.{\alp{n(2p-1)+p}}.$}
\put(176,57){$\ncddeof{}.{}.
     {}.{\alp{n(2p+1)+p}}.$}
\put(245,27){$\Downarrow$}
\put(142,5){$\eddanid{\bet{0}}.{\bet{1}}.{\bet{2}}.
                {\bet{n-1}}.{\bet{n}}.$}
\end{picture}
\caption{$a_{2(2np+n+p)}^{(2)}\rightarrow$ \atnt($a_{2p}^{(2)}$)}
\label{fig:aat}
\end{figure}
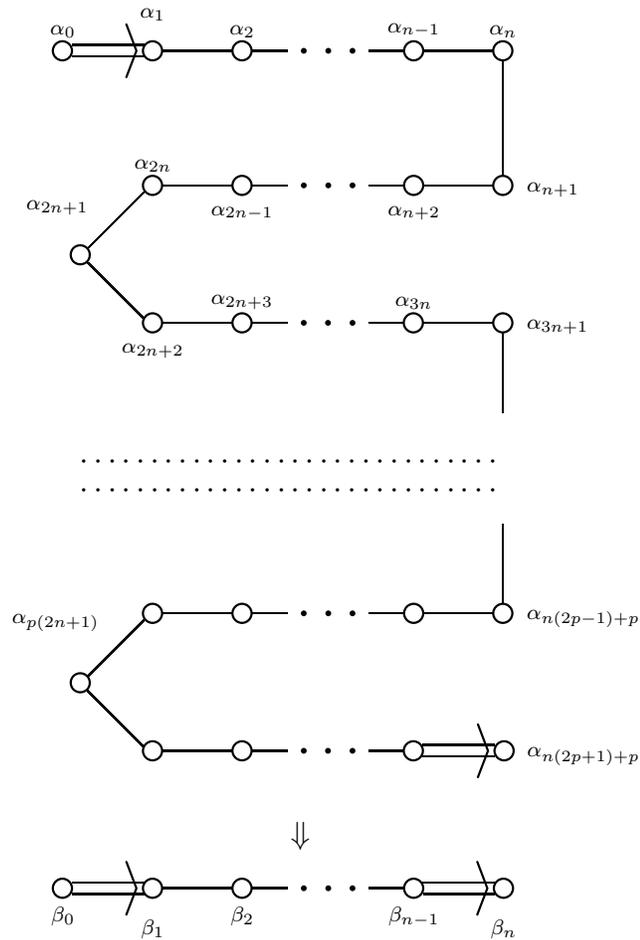

\setcounter{equation}{0}
The series \atnt\ has only one reduction viz it reduces to a member of
the same series. The reduction is depicted in Fig.\ts\ref{fig:aat} for
$a_{2(2np+n+p)}^{(2)}$ going to \atnt($a_{2p}^{(2)}$). In this case
again we see that at one edge the neighbouring vertices are identified
and at the opposite edge not. And the other difference from the
earlier Figs. \ref{fig:dat} and \ref{fig:cat} is the series begins at one
edge and ends at the other.

\section{\dno, \bno, and \aont\ theories}
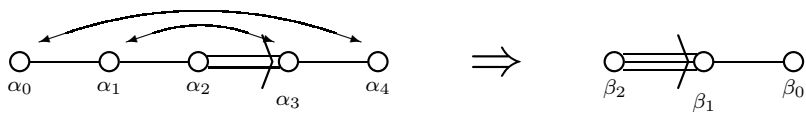
\begin{figure}
\begin{picture}(400,555)
\put(5,545){i) $d_5^{(1)}~\rightarrow~c_2^{(1)}$}
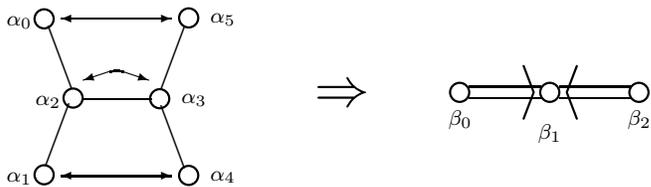
\put(20,490){$\edddndf{\alp{0}}.{\alp{2}}.{\alp{1}}.{\alp{3}}.{\alp{4}}.
{\alp{5}}.\quad\Rightarrow\quad \eddciii{\bet{0}}.{\bet{1}}.{\bet{2}}.$}
\put(54,503){\vector(-3,-1){10}}
\put(58.8,503){\vector(3,-1){10}}
\qbezier(54,503)(56.4,503.8)(58.8,503)
\put(56.4,523){\vector(-1,0){21}}
\put(56.4,523){\vector(1,0){21}}
\put(56.4,463.5){\vector(-1,0){21}}
\put(56.4,463.5){\vector(1,0){21}}
\put(5,435){ii) $d_4^{(1)}~\rightarrow~g_2^{(1)}$}
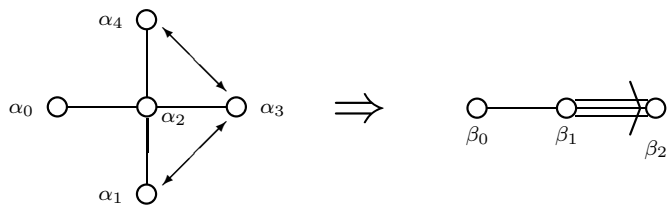
\put(5,380){$\edddiv{\alp{0}}.{\alp{4}}.{\alp{3}}.{\alp{1}}.
\quad\Rightarrow\quad \eddgid{\bet{0}}.{\bet{1}}.{\bet{2}}.$}
\put(61,380){$\scriptstyle \alp{2}$}
\put(73.5,403.5){\vector(1,-1){12}}
\put(73.5,403.5){\vector(-1,1){12}}
\put(73.5,368){\vector(-1,-1){12}}
\put(73.5,368){\vector(1,1){12}}
\put(5,325){iii) $a_{2n+1}^{(2)}~\rightarrow~a_{2n}^{(2)}$}
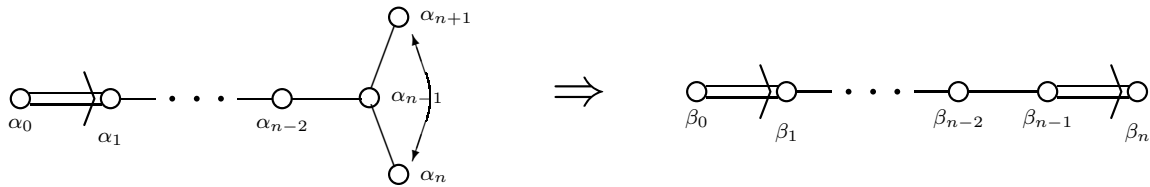
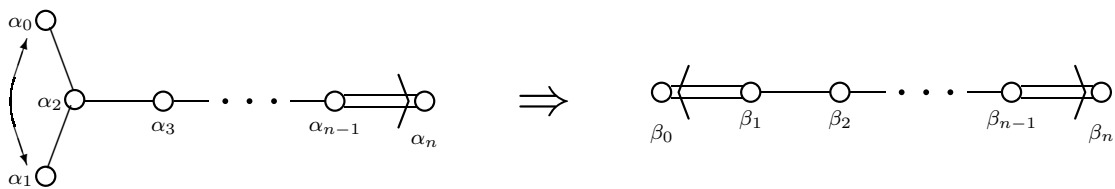
\put(5,270){$\eddanod{\alp{0}}.{\alp{1}}.{\alp{n-2}}.
            {\alp{n-1}}.{\alp{n}}.{\alp{n+1}}.\quad\Rightarrow\quad
\eddanidr{\bet{0}}.{\bet{1}}.{\bet{n-2}}.{\bet{n-1}}.{\bet{n}}.$}
\put(175,282){\vector(-1,3){5}}
\put(175,264){\vector(-1,-3){5}}
\qbezier(175,282)(178,273)(175,264)
\put(5,215){iv) $b_{n}^{(1)}~\rightarrow~d_{n}^{(2)}$}
\put(15,160){$\eddbnd{\alp{0}}.{\alp{2}}.{\alp{1}}.{\alp{3}}.{\alp{n-1}}.
{\alp{n}}.\quad\Rightarrow\quad
\edddniid{\bet{0}}.{\bet{1}}.{\bet{2}}.{\bet{n-1}}.{\bet{n}}.$}
\put(12,174){\vector(1,3){5}}
\put(12,156){\vector(1,-3){5}}
\qbezier(12,174)(9,165)(12,156)
\put(5,105){v) $f_{4}^{(1)}~\rightarrow~d_{4}^{(3)}$}
\put(21,50){$\eddfid{\alp{0}}.{\alp{1}}.{\alp{2}}.{\alp{3}}.{\alp{4}}.
\quad\Rightarrow\quad \eddgiid{\bet{2}}.{\bet{1}}.{\bet{0}}.$}
\put(88,66){\vector(-3,-1){10}}
\put(124,66){\vector(3,-1){10}}
\qbezier(88,66)(106,72)(124,66)
\put(55,66){\vector(-3,-1){10}}
\put(157,66){\vector(3,-1){10}}
\qbezier(55,66)(106,83)(157,66)
\end{picture}
\caption{Reductions of $d_5^{(1)}$, $d_4^{(1)}$, $a_{2n+1}^{(1)}$,
        \bno, \ffo}\label{fig:dab}
\end{figure}

\setcounter{equation}{0}
 Most of the reductions and foldings of \dno\ series are known for a long
time and discussed in many places \cite{BCDSc,OT,Sas} earlier. Here for
completeness sake we give them in Fig.\ts\ref{fig:dno}. The new one we
have added in this figure is the last one, where $d_{2n+1}^{(1)}$
reduces to \bno\ theory. In all these cases we identify vertices as
shown in the Fig.\ts\ref{fig:dno} (with external arrows).
There are other particular ones we have added,
for example Fig.\ts\ref{fig:dab}, where $d_5^{(1)}$ reduces to $c_2^{(1)}$.
Another interesting feature is that one can obtain a reduction through
multiple steps. Let us consider the reduction of $d_4^{(1)}$ to $a_2^{(2)}$.
This can be achieved by identifying four outer vertices of the diagram
at one step. The same thing can be achieved in the following way. First
only two of the outer vertices are identified to obtain $b_3^{(1)}$. Now
the roots $\alp1$ and $\alp3$ of $b_3^{(1)}$ are identified to obtain
the $g_2^{(1)}$ theory. Note that in this case the two vertices
identified have different `weights' so one has to take the weighted
average for the construction of the new roots in this case as
explained in the section 3, (\ref{wetav}). In the final step we identify
the roots $\alp0$ and $\alp2$ of $g_2^{(1)}$ theory to get $a_2^{(2)}$. Here
we obtain the final theory in three steps. The same thing could have
been achieved in two steps viz  $d_4^{(1)}\rightarrow
d_2^{(2)}\rightarrow a_2^{(2)}$. The reductions of \bno\ and \aont\ are
very limited and are given in Fig. \ref{fig:dab}.

\section{Exceptional series}
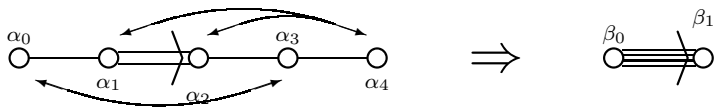
\begin{figure}
\begin{picture}(400,550)
\put(5,540){i) $e_{7}^{(1)}~\rightarrow~e_{6}^{(2)}$}
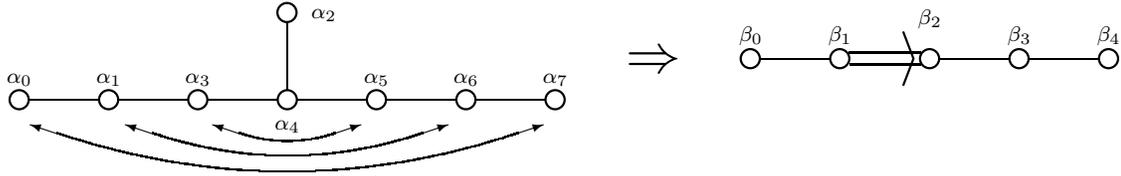
\put(20,480){$\eddeii{\alp{1}}.{\alp{2}}.{\alp{3}}.{\alp{4}}.{\alp{5}}.
{\alp{6}}.{\alp{7}}.{\alp{0}}.~\Rightarrow~
\eddfiiu{\bet{0}}.{\bet{1}}.{\bet{2}}.{\bet{3}}.{\bet{4}}.$}
\put(225,457){\vector(3,1){10}}
\put(51,457){\vector(-3,1){10}}
\qbezier(51,457)(138,428)(225,457)
\put(189,457){\vector(3,1){10}}
\put(87,457){\vector(-3,1){10}}
\qbezier(189,457)(138,440)(87,457)
\put(156,457){\vector(3,1){10}}
\put(120,457){\vector(-3,1){10}}
\qbezier(120,457)(138,451)(156,457)
\put(5,390){ii) $e_{6}^{(1)}~\rightarrow~f_{4}^{(1)}$}
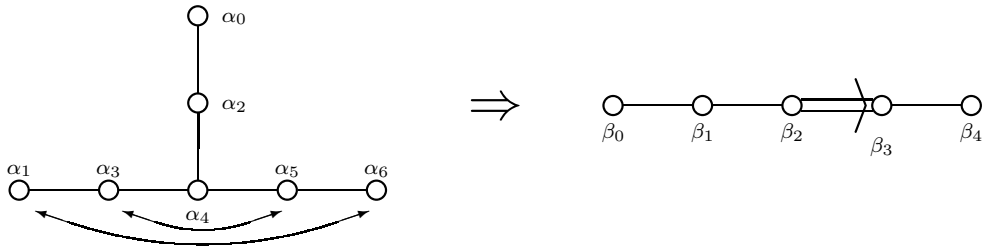
\put(20,320){$\eddei{\alp{1}}.{\alp{2}}.{\alp{3}}.{\alp{4}}.{\alp{5}}.
{\alp{6}}.{\alp{0}}.\quad\Rightarrow\quad
\eddfid{\bet{0}}.{\bet{1}}.{\bet{2}}.{\bet{3}}.{\bet{4}}.$}
\put(88,281){\vector(-3,1){12}}
\put(124,281){\vector(3,1){12}}
\qbezier(88,281)(106,275)(124,281)
\put(55,281){\vector(-3,1){12}}
\put(157,281){\vector(3,1){12}}
\qbezier(55,281)(106,264)(157,281)
\put(5,230){iii) $e_{6}^{(1)}~\rightarrow~d_{4}^{(3)}$}
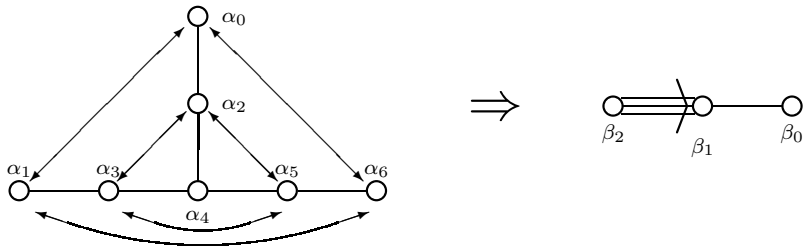
\put(20,160){$\eddei{\alp{1}}.{\alp{2}}.{\alp{3}}.{\alp{4}}.{\alp{5}}.
{\alp{6}}.{\alp{0}}.\quad\Rightarrow\quad
\eddgiid{\bet{2}}.{\bet{1}}.{\bet{0}}.$}
\put(88,121){\vector(-3,1){12}}
\put(124,121){\vector(3,1){12}}
\qbezier(88,121)(106,115)(124,121)
\put(55,121){\vector(-3,1){12}}
\put(157,121){\vector(3,1){12}}
\qbezier(55,121)(106,104)(157,121)
\put(121.5,150){\vector(-1,1){13}}
\put(121.5,150){\vector(1,-1){13}}
\put(87,150){\vector(1,1){13}}
\put(87,150){\vector(-1,-1){13}}
\put(138,166){\vector(-1,1){29}}
\put(138,166){\vector(1,-1){29}}
\put(70,166){\vector(1,1){29}}
\put(70,166){\vector(-1,-1){29}}
\put(5,70){iv) $e_{6}^{(2)}~\rightarrow~a_{2}^{(2)}$}
\put(20,40){$\eddfiid{\alp{0}}.{\alp{1}}.{\alp{2}}.{\alp{3}}.{\alp{4}}.
\quad\Rightarrow\quad \datwot{\bet0}.{\bet1}.$}
\put(87,58){\vector(-3,-1){12}}
\put(120,58){\vector(-3,-1){12}}
\put(124,33){\vector(3,1){12}}
\qbezier(120,58)(138,64)(156,58)
\qbezier(87,58)(121.5,69.5)(156,58)
\put(55,33){\vector(-3,1){12}}
\put(156,58){\vector(3,-1){12}}
\qbezier(55,33)(89.5,21.5)(124,33)
\end{picture}
\caption{Exceptional reductions}\label{fig:ecep}
\end{figure}

All the reductions of the exceptional theories were discussed in the
papers \cite{BCDSc,OT,Sas}. Here we only add one more step in the
reduction of \esio\ to \dfth\ ie. \esio $\rightarrow$ \ffo $\rightarrow$
\dfth. The vertices to be identified in this case are shown in Fig
\ref{fig:ecep} through arrows. As remarked in a previous paper \cite{Sas},
one can reduce directly $e_7^{(1)}\to a_2^{(2)}$, by keeping (or
projecting onto) the
$({\rm mass})^2=6m^2$ particle only. The same reduction can be
achieved in two steps via $e_6^{(2)}$, that is $e_7^{(1)}\rightarrow
e_6^{(2)}\rightarrow a_2^{(2)}$(first and last diagrams of Fig.
\ref{fig:ecep}).

\section{Summary and discussion}
The reductions of affine Toda field theories are reported
systematically and comprehensively by adopting simple graphical
representations indicating the vertices of the Dynkin diagrams to be
identified (folded).
Many new reductions are reported and they are expected to play
important roles: for example in understanding the `soliton'
S-matrices and their fusions associated with the `imaginary' coupling
theories, in the representation theory of affine algebras, etc.


\pagebreak
\begin{figure}
\begin{picture}(300,100)
\put(111,83){$\lronit{\alp{0}}.$}
\put(142,96){$\ncdfaur{\alp{1}}.{\alp{2}}.{\alp{n-2}}.
     {\alp{n-1}}.{\alp{n}}.$}
\put(142,45){$\ddanddf{\alp{2n-1}}.{\alp{2n-2}}.
     {\alp{n+2}}.{\alp{n+1}}.$}
\put(222,27){$\Downarrow$}
\put(109,5){$\ddcnds{\bet{0}}.{\bet{1}}.{\bet{2}}.{\bet{n-2}}.
     {\bet{n-1}}.{\bet{n}}.$}
\end{picture}
\caption{$a_{2n-1}^{(1)}\rightarrow c_n^{(1)}$}
\label{aotc}
\end{figure}
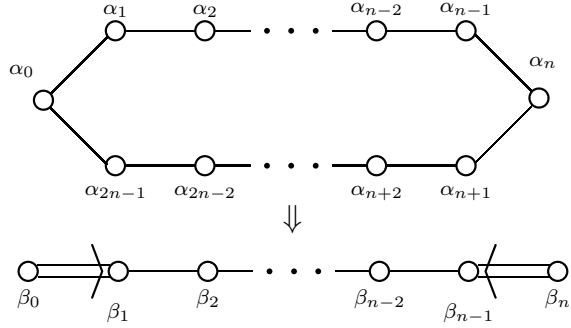

\begin{figure}
\begin{picture}(300,100)
\put(120,80){$\ncddlr{\alp{0}}.{\alp{1}}.{\alp{2}}.
         {\alp{n-1}}.{\alp{n}}.$}
\put(120,46){$\ncdulr{\alp{2n+1}}.{\alp{2n}}.{\alp{2n-1}}.
     {\alp{n+2}}.{\alp{n+1}}.$}
\put(218,27){$\Downarrow$}
\put(120,5){$\edddniid{\bet{0}}.{\bet{1}}.{\bet{2}}.
         {\bet{n-1}}.{\bet{n}}.$}
\end{picture}
\caption{$a_{2n+1}^{(1)}\rightarrow d_{n+1}^{(2)}$}
\label{aotd}
\end{figure}

\begin{figure}
\begin{picture}(300,120)
\put(111,83){$\lronit{\alp{0}}.$}
\put(142,96){$\ncddrdu{\alp{1}}.{\alp{2}}.{\alp{n-2}}.
     {\alp{n-1}}.{\alp{n}}.$}
\put(142,45){$\ncddrud{\alp{2n}}.{\alp{2n-1}}.{\alp{2n-2}}.
     {\alp{n+2}}.{\alp{n+1}}.$}
\put(222,27){$\Downarrow$}
\put(109,5){$\datwon{\bet{0}}.{\bet{1}}.{\bet{2}}.{\bet{n-2}}.
     {\bet{n-1}}.{\bet{n}}.$}
\end{picture}
\caption{$a_{2n}^{(1)}\rightarrow a_{2n}^{(2)}$}
\label{aoat}
\end{figure}
\clearpage

\begin{figure}
\begin{picture}(300,225)
\put(95,215){a)~$q$~even}
\put(315,215){b)~$q$~odd}
\put(10,185){$\ncddd{\alp{0}}.{\alp{1}}.{\alp{2}}.
     {\alp{p-2}}.{\alp{p-1}}.$}
\put(10,151.5){$\ncanur{\alp{2p-1}}.{\alp{2p-2}}.{\alp{2p-3}}.
     {\alp{p+1}}.{\alp{p}}.$}
\put(30,129.5){$\cdots\cdots\cdots\cdots\cdots\cdots\cdots\cdots%
       \cdots\cdots$}
\put(30,116){$\cdots\cdots\cdots\cdots\cdots\cdots\cdots\cdots%
        \cdots\cdots$}
\put(30,102.5){$\cdots\cdots\cdots\cdots\cdots\cdots\cdots\cdots%
         \cdots\cdots$}
\put(10,78.5){$\ncandr{\alp{p(q-2)}}.{}.{}.
     {}.{}.$}
\put(10,45){$\ncddu{\alp{pq-1}}.{\alp{pq-2}}.{\alp{pq-3}}.
     {}.{\alp{p(q-1)}}.$}
\put(100,27){$\Downarrow$}
\put(10,5){$\edddniid{\bet{0}}.{\bet{1}}.{\bet{2}}.
     {\bet{p-2}}.{\bet{p-1}}.$}
\put(230,185){$\ncddd{\alp{0}}.{\alp{1}}.{\alp{2}}.
     {\alp{p-2}}.{\alp{p-1}}.$}
\put(230,151.5){$\ncanur{\alp{2p-1}}.{\alp{2p-2}}.{\alp{2p-3}}.
     {\alp{p+1}}.{\alp{p}}.$}
\put(250,129.5){$\cdots\cdots\cdots\cdots\cdots\cdots\cdots%
      \cdots\cdots\cdots$}
\put(250,116){$\cdots\cdots\cdots\cdots\cdots\cdots\cdots%
       \cdots\cdots\cdots$}
\put(250,102.5){$\cdots\cdots\cdots\cdots\cdots\cdots\cdots%
        \cdots\cdots\cdots$}
\put(230,78.5){$\ncanur{}.{}.{}.{}.{\alp{p(q-2)}}.$}
\put(230,45){$\ncdduo{\alp{p(q-1)}}.{}.
{}.{\alp{pq-2}}.{\alp{pq-1}}.$}
\put(320,27){$\Downarrow$}
\put(230,5){$\edddniid{\bet{0}}.{\bet{1}}.{\bet{2}}.
     {\bet{p-2}}.{\bet{p-1}}.$}
\end{picture}
\caption{$d_{pq}^{(2)} \rightarrow d_{p}^{(2)}(d_{q}^{(2)})$}
\label{fig:dtd}
\end{figure}

\begin{figure}
\begin{picture}(350,275)(-60,0)
\put(40,230){$d_{2q}^{(2)}\rightarrow~ a_1^{(1)}$}
\put(5,215){a)~$q$~even}
\put(75,215){b)~$q$~odd}
\put(10,185){$\ncdddt{\alp{0}}.{\alp{1}}.$}
\put(10,151.5){$\ncanurt{\alp{3}}.{\alp{2}}.$}
\put(30,129.5){$\cdots\cdots$}
\put(30,116){$\cdots\cdots$}
\put(30,102.5){$\cdots\cdots$}
\put(10,78.5){$\ncandrt{\alp{2q-4}}.{\alp{2q-3}}.$}
\put(10,45){$\ncddut{\alp{2q-1}}.{\alp{2q-2}}.$}
\put(43,27){$\Downarrow$}
\put(10,5){$\daone{\bet{0}}.{\bet{1}}.$}
\put(80,185){$\ncdddt{\alp{0}}.{\alp{1}}.$}
\put(80,151.5){$\ncanurt{\alp{3}}.{\alp{2}}.$}
\put(100,129.5){$\cdots\cdots$}
\put(100,116){$\cdots\cdots$}
\put(100,102.5){$\cdots\cdots$}
\put(80,78.5){$\ncanurt{\alp{2q-3}}.{\alp{2q-4}}.$}
\put(80,45){$\ncdduot{\alp{2q-2}}.{\alp{2q-1}}.$}
\put(110,27){$\Downarrow$}
\put(80,5){$\daone{\bet{0}}.{\bet{1}}.$}
\put(210,265){$c_{q}^{(1)}\rightarrow~ a_1^{(1)}$}
\put(175,250){a)~$q$~even}
\put(245,250){b)~$q$~odd}
\put(179,241){$\laronit{\alp{0}}.{\alp{1}}.$}
\put(180,190){$\lronit{\alp{2}}.$}
\put(195,146){$\cdots\cdots$}
\put(195,132.5){$\cdots\cdots$}
\put(180,108.5){$\lronit{\alp{q-2}}.$}
\put(179,89.5){$\rarsnit{\alp{q-1}}.{\alp{q}}.$}
\put(213,27){$\Downarrow$}
\put(180,5){$\daone{\bet{0}}.{\bet{1}}.$}
\put(249,241){$\laronit{\alp{0}}.{\alp{1}}.$}
\put(250,190){$\lronit{\alp{2}}.$}
\put(270,146){$\cdots\cdots$}
\put(270,132.5){$\cdots\cdots$}
\put(245.5,125.5){$\rronit{\alp{q-2}}.$}
\put(239,87){$\raronit{\alp{q-1}}.{\alp{q}}.$}
\put(283,27){$\Downarrow$}
\put(250,5){$\daone{\bet{0}}.{\bet{1}}.$}
\end{picture}
\caption{$d_{2q}^{(2)}(c_{q}^{(1)}) \rightarrow a_{1}^{(1)}$}
\label{fig:dcao}
\end{figure}
\clearpage

\clearpage

\clearpage

\clearpage

\begin{figure}
\begin{picture}(300,355)
\put(142,352){$\ncddcd{\alp{0}}.{\alp{1}}.{\alp{2}}.
         {\alp{n-1}}.{\alp{n}}.$}
\put(145,288){$\ncdfalu{\alp{2n+1}}.{\alp{2n}}.{\alp{2n-1}}.
     {\alp{n+2}}.{\alp{n+1}}.$}
\put(176,249){$\ddanufd{\alp{2n+2}}.
     {\alp{2n+3}}.{\alp{3n}}.{\alp{3n+1}}.$}
\put(165,204){$\cdots\cdots\cdots\cdots\cdots\cdots\cdots\cdots%
         \cdots\cdots$}
\put(165,193){$\cdots\cdots\cdots\cdots\cdots\cdots\cdots%
             \cdots\cdots\cdots$}
\put(145,135){$\ncdfalu{\alp{(p-1)(2n+1)}}.{}.{}.
     {}.{\alp{(p-1)(2n+1)-n}}.$}
\put(176,96){$\ddanufd{}.{}.
     {}.{\alp{p(2n+1)-n-1}}.$}
\put(142,45){$\ncddcu{\alp{p(2n+1)}}.{}.{}.
     {}.{\alp{p(2n+1)-n}}.$}
\put(245,27){$\Downarrow$}
\put(142,5){$\eddanid{\bet{0}}.{\bet{1}}.{\bet{2}}.
           {\bet{n-1}}.{\bet{n}}.$}
\end{picture}
\caption{$c_{p(2n+1)}^{(1)}\rightarrow$ \atnt} \label{fig:cat}
\end{figure}
\clearpage

\clearpage

\begin{figure}
\begin{picture}(400,555)
\put(5,545){i) $d_{n+1}^{(1)}\rightarrow$ \bno}
\put(20,490){$\edddnd{\alp{0}}.{\alp{2}}.{\alp{1}}.{\alp{3}}.{\alp{n-2}}.
{\alp{n-1}}.{\alp{n}}.{\alp{n+1}}.~\Rightarrow~
\eddbnd{\bet{0}}.{\bet{2}}.{\bet{1}}.{\bet{3}}.{\bet{n-1}}.{\bet{n}}.$}
\put(192,502){\vector(-1,3){5}}
\put(192,484){\vector(-1,-3){5}}
\qbezier(192,502)(195,493)(192,484)
\put(5,435){ii) $d_{n+2}^{(1)}\rightarrow~ d_{n+1}^{(2)}$}
\put(20,380){$\edddnd{\alp{0}}.{\alp{2}}.{\alp{1}}.{\alp{3}}.{\alp{n-1}}.
{\alp{n}}.{\alp{n+1}}.{\alp{n+2}}.~\Rightarrow~
\edddniid{\bet{0}}.{\bet{1}}.{\bet{2}}.{\bet{n-1}}.{\bet{n}}.$}
\put(192,392){\vector(-1,3){5}}
\put(192,374){\vector(-1,-3){5}}
\qbezier(192,392)(195,383)(192,374)
\put(20,392){\vector(1,3){5}}
\put(20,374){\vector(1,-3){5}}
\qbezier(20,392)(17,383)(20,374)
\put(5,325){iii) $d_{2n}^{(1)}\rightarrow$ \aont}
\put(5,270){$\edddnds{\alp{0}}.{\alp{2}}.{\alp{1}}.{\alp{3}}.{\alp{n}}.
{\alp{2n-3}}.{\alp{2n-2}}.
{\alp{2n-1}}.{\alp{2n}}.~\Rightarrow~
\eddanod{\bet{0}}.{\bet{1}}.{\bet{n-3}}.{\bet{n-2}}.
           {\bet{n-1}}.{\bet{n}}.$}
\put(73,283){\vector(-3,-1){10}}
\put(154,283){\vector(3,-1){10}}
\qbezier(73,283)(113.5,296.5)(154,283)
\put(40,283){\vector(-3,-1){10}}
\put(187,283){\vector(3,-1){10}}
\qbezier(40,283)(113.5,307.5)(187,283)
\put(197,303){\vector(1,0){7}}
\put(30,303){\vector(-1,0){7}}
\put(30,303){\line(1,0){167}}
\put(30,243.5){\vector(-1,0){7}}
\put(197,243.5){\vector(1,0){7}}
\put(30,243.5){\line(1,0){167}}
\put(5,215){iv) $d_{2n+2}^{(1)}\rightarrow$ \atnt}
\put(5,160){$\edddnds{\alp{0}}.{\alp{2}}.{\alp{1}}.{\alp{3}}.{\alp{n}}.
{\alp{2n-1}}.{\alp{2n}}.{\alp{2n+1}}.{\alp{2n+2}}.~~
\eddanidr{\bet{0}}.{\bet{1}}.{\bet{n-2}}.{\bet{n-1}}.{\bet{n}}.$}
\put(73,173){\vector(-3,-1){10}}
\put(154,173){\vector(3,-1){10}}
\qbezier(73,173)(113.5,186.5)(154,173)
\put(40,173){\vector(-3,-1){10}}
\put(187,173){\vector(3,-1){10}}
\qbezier(40,173)(113.5,197.5)(187,173)
\put(197,193){\vector(1,0){7}}
\put(30,193){\vector(-1,0){7}}
\put(30,193){\line(1,0){167}}
\put(30,133.5){\vector(-1,0){7}}
\put(197,133.5){\vector(1,0){7}}
\put(30,133.5){\line(1,0){167}}
\put(222,172){\vector(-1,3){5}}
\put(222,154){\vector(-1,-3){5}}
\qbezier(222,172)(225,163)(222,154)
\put(230,163){$\Rightarrow$}
\put(5,105){v) $d_{2n+1}^{(1)}\rightarrow$ \bno}
\put(20,50){$\edddnd{\alp{0}}.{\alp{2}}.{\alp{1}}.{\alp{3}}.{\alp{n-2}}.
{\alp{n-1}}.{\alp{n}}.{\alp{n+1}}.~\Rightarrow~
\eddbnd{\bet{0}}.{\bet{2}}.{\bet{1}}.{\bet{3}}.{\bet{n-1}}.{\bet{n}}.$}
\put(88,63){\vector(-3,-1){10}}
\put(124,63){\vector(3,-1){10}}
\qbezier(88,63)(106,69)(124,63)
\put(55,63){\vector(-3,-1){10}}
\put(157,63){\vector(3,-1){10}}
\qbezier(55,63)(106,80)(157,63)
\put(167,83){\vector(1,0){7}}
\put(45,83){\vector(-1,0){7}}
\put(45,83){\line(1,0){122}}
\put(45,23.5){\vector(-1,0){7}}
\put(167,23.5){\vector(1,0){7}}
\put(45,23.5){\line(1,0){122}}
\end{picture}
\caption{Reductions of \dno series}\label{fig:dno}
\end{figure}
\clearpage

\clearpage

\begin{figure}
\begin{picture}(300,350)
\put(5,335){i) $b_{3}^{(1)}~\rightarrow~g_{2}^{(1)}$}
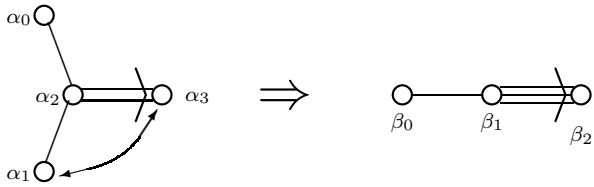
\put(20,270){$\eddbndt{\alp{0}}.{\alp{2}}.{\alp{1}}.{\alp{3}}.
\quad\Rightarrow\quad \eddgid{\bet0}.{\bet1}.{\bet2}.$}
\put(65,259){\vector(2,3){7}}
\put(45,247){\vector(-4,-1){10}}
\qbezier(45,247)(57.5,249.5)(65,259)
\put(5,205){ii) $g_{2}^{(1)}~\rightarrow~a_{2}^{(2)}$}
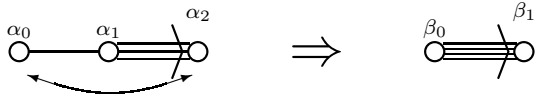
\put(20,160){$\eddgiu{\alp{0}}.{\alp{1}}.{\alp{2}}.
\quad\Rightarrow\quad \datwot{\bet0}.{\bet1}.$}
\put(50,153){\vector(-3,1){10}}
\put(92,153){\vector(3,1){10}}
\qbezier(50,153)(71,146)(92,153)
\put(5,115){iii) $b_{3}^{(1)}~\rightarrow~a_{2}^{(2)}$}
\put(20,50){$\eddbndt{\alp{0}}.{\alp{2}}.{\alp{1}}.{\alp{3}}.
\quad\Rightarrow\quad \datwot{\bet0}.{\bet1}.$}
\put(65,71.5){\vector(2,-3){7}}
\put(45,83.5){\vector(-4,1){10}}
\qbezier(45,83.5)(57.5,81)(65,71.5)
\put(65,39){\vector(2,3){7}}
\put(45,27){\vector(-4,-1){10}}
\qbezier(45,27)(57.5,29.5)(65,39)
\end{picture}
\caption{Reductions of $b_3^{(1)}$ and \gto}\label{fig:bng}
\end{figure}
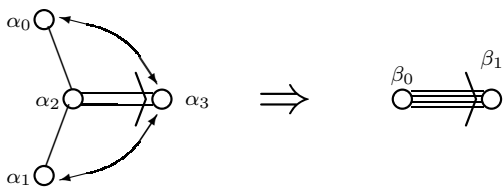
\clearpage

\end{document}